\documentclass[draftcls,onecolumn,11pt]{IEEEtran}

\usepackage{cite}      % Written by Donald Arseneau
\usepackage{graphicx}  % Written by David Carlisle and Sebastian Rahtz
\usepackage{psfrag}    % Written by Craig Barratt, Michael C. Grant,
\usepackage{subfigure} % Written by Steven Douglas Cochran
\usepackage{url}       % Written by Donald Arseneau
\usepackage{amssymb,amsmath,amscd,amsfonts,amstext}
\usepackage{array}
\usepackage{multirow}

\newcommand{\RR}{{\mathbb{R}}}

\newcommand{\CC}{{\mathbb{C}}}
\newcommand{\EE}{{\mathbb{E}}}
\newcommand{\tr}{{\mathrm{tr}}}
\newcommand{\bs}{\boldsymbol}
\newcommand{\diag}{{\mathrm{diag}}}
\newcommand{\overcirc}{\overset{\circ}}
\newtheorem{lemma}{Lemma}

\newtheorem{proposition}{Proposition}
\newtheorem{theorem}{Theorem}

\newtheorem{remark}{Remark}
\def\adots{
  \mathinner{\mkern1mu\raise1pt\hbox{.}\mkern2mu\raise4pt\hbox{.}
  \mkern2mu\raise7pt\vbox{\kern7pt\hbox{.}}\mkern1mu}}
\def\build#1_#2^#3{\mathrel{
\mathop{\kern 0pt#1}\limits_{#2}^{#3}}}
\IEEEoverridecommandlockouts 

\begin{document}

% paper title
\title{A New Approach for Capacity Analysis of 
Large Dimensional Multi-Antenna Channels} 

\author{W. Hachem$^{(*)}$, O. Khorunzhiy, Ph. Loubaton, J. Najim and L. Pastur
\thanks{This work was partially supported by the ``Fonds National de la Science'' via the ACI program ``Nouvelles Interfaces des Math\'ematiques'', project MALCOM number 205\ .}% <-this % stops a space
\thanks{W. Hachem and J. Najim are with
CNRS and ENST (UMR 5141), Paris, France.
        {\tt \{hachem,najim\}@enst.fr}\ ,}
\thanks{O. Khorunzhiy is with Equipe "Probabilit\'es-Statistiques"
Universit\'e de Versailles - Saint-Quentin, 
France {\tt khorunjy@math.uvsq.fr}\ ,}  
\thanks{P. Loubaton is with IGM LabInfo, UMR 8049, Institut Gaspard Monge,
Universit\'e de Marne La Vall\'ee, France. 
        {\tt loubaton@univ-mlv.fr}\ ,}%
\thanks{L. Pastur is with Kharkiv National University - Institute for Low Temperature Physics
 - Kharkiv, Ukraine {\tt lpastur@flint.ilt.kharkov.ua}\ .
}
\thanks{(*) Corresponding author. 
}}
\date{15 December 2006}

% The paper headers
\markboth{submitted to \emph{IEEE Transactions on Information Theory}}
{Hachem \emph{et.al.}}

% make the title area
\maketitle

\begin{abstract}
  This paper adresses the behaviour of the mutual information of
  correlated MIMO Rayleigh channels when the numbers of transmit and
  receive antennas converge to $+\infty$ at the same rate. Using a
  new and simple approach based on 
  Poincar\'e-Nash inequality and on an integration by parts formula, it
  is rigorously established that the mutual information converges to a
  Gaussian random variable whose mean and variance are evaluated. These
  results confirm previous evaluations based on the powerful but non
  rigorous replica method. It is believed that the tools that are used
  in this paper are simple, robust, and of interest for the
  communications engineering community.
\end{abstract}


\section{Introduction}
\label{sec-intro}

It is widely known that high spectral efficiencies are attained when
multiple antennas are used at both the transmitter and the receiver of
a wireless communication system. Indeed, due to the mobility and to
the presence of a large number of reflected and scattered signal
paths, the elements of the $N \times n$ Multiple Input Multiple Output
(MIMO) channel matrix with $N$ antennas at the receiver's site and $n$
antennas at the transmitter's are often modeled as random variables.
Assuming a random model for this matrix, Telatar realized in the
mid-nineties that Shannon's capacity of such channels increases at the
rate of $\min(N,n)$ for a fixed transmission power \cite{tel-95}. A
result of the same nature can be found in the work of Foschini and
Gans \cite{fos-gan-98}.  The authors of \cite{tel-95} and
\cite{fos-gan-98} assumed that the elements of the channel matrix
${\bf G}$ are centered, independent and identically distributed
(i.i.d.)  elements. In this context, a well known result in Random
Matrix Theory (RMT) \cite{mar-pas-ms67} says that the eigenvalue
distribution of the Gram matrix ${\bf GG}^*$ where ${\bf G}^*$ is the
Hermitian adjoint of ${\bf G}$ converges to a deterministic
probability distribution as $n$ goes to infinity and $N/n$ converges
to a constant $c > 0$.  Denote by $I(\rho) = \log\det\left(
  \frac{\rho}{n} {\bf GG}^* + {\bf I}_N \right)$ the capacity of
channel ${\bf G}$ for a Signal to Noise Ratio at a receiver antenna
equal to $\rho / n$.  One consequence of \cite{mar-pas-ms67} is that
the capacity per transmit antenna $I(\rho)/n$, being an integral of a
$\log$ function with respect to the empirical eigenvalue distribution
of ${\bf GG}^*$, converges to a constant.  This fact already observed
in \cite{tel-95} sustains the assertion of the linear increase of
capacity with the number of antennas.  In addition, this convergence
proves to be sufficiently fast. As a matter of fact, the asymptotic
results predicted by the RMT remain relevant for
systems with a moderate number of antennas. \\
The next step was to apply this theory to channel models that include
a correlation between paths (or entries of ${\bf G}$).  One of the
main purposes of this generalization is to better understand the
impact of these correlations on Shannon's mutual information. Let us
cite in this context the contributions \cite{mue-it02},
\cite{chu-tse-kah-val-it02}, \cite{mes-fon-pag-jsac03},
\cite{mou-sim-sen-it03} and \cite{tul-loz-ver-it05}, all devoted to
the study of the mutual information in the case where the elements of
channel's matrix are centered and correlated random variables.  In
\cite{hac-lou-naj-(sub)aap05}, a deterministic equivalent is computed
under broad conditions for the capacity based on Rice channels modeled
by non-centered matrices with independent but not identically
distributed random variables. The link between matrices with
correlated entries and matrices with independent entries and a
variance profile is
studied in \cite{HLN05}.\\

One of the most popular correlated channel models used for these
capacity evaluations is the so-called Kronecker model ${\bf G} = {\bs
  \Psi} {\bf W} \widetilde{\bs \Psi}$ where ${\bf W}$ is a $N \times
n$ matrix with Gaussian centered i.i.d. entries, and ${\bs \Psi}$ and
$\widetilde{\bs \Psi}$ are $N \times N$ and $n \times n$ matrices that
capture the path correlations at the receiver and at the transmitter
sides respectively \cite{shi-fos-gan-kah-tcom00},
\cite{ker-sch-perd-mog-fre-jsac02}.  This model has been studied by
Chuah et. al. in \cite{chu-tse-kah-val-it02}.  With some assumptions
on matrices ${\bs \Psi}$ and $\widetilde{\bs \Psi}$, these authors
showed that $I (\rho) / n$ converges to a deterministic quantity
defined as the fixed point of an integral equation.  Later on, Tulino
et. al.  \cite{tul-loz-ver-it05} obtained the limit of $I (\rho) / n$
for a correlation model more general than the Kronecker model.  Both
these works rely on a result of Girko describing the eigenvalue
distribution of the Gram matrix associated with a matrix with
independent but non necessarily identically distributed entries, a
close model as we shall see in a moment. \\
In \cite{mou-sim-sen-it03}, Moustakas et. al. studied the mutual
information for the Kronecker model by using the so-called replica
method. They found an approximation $V(\rho)$ of $\EE\left[ I(\rho)
\right]$ accurate to the order $1/n$ in the large $n$ regime.  Using
this same method, they also showed that the variance of $I(\rho) -
V(\rho)$ is of
order one and were able to derive this variance for large $n$. \\
Although the replica technique is powerful and has a wide range of
applications, the rigorous justification of some of its parts remains
to be done. In this paper, we propose a new method to study the
convergence of $\mathbb{E} I(\rho)$ and the fluctuations of $I(\rho)$.
Beside recovering the results in \cite{mou-sim-sen-it03}, we establish
the Central Limit Theorem (CLT) for $I(\rho) - V(\rho)$. The practical
interest of such a result is of importance since the CLT leads to an
evaluation of the outage probability, i.e. the probability that
$I(\rho)$ lies beneath a given threshold, by means of the Gaussian
approximation. Many other works have been devoted to CLT for random
matrices. Close to our present article are \cite{AndZei06},
\cite{BaiSil04}, \cite{BouKho98}.

In this article, we also would like to advocate the method used to
establish both the approximation of $I(\rho)$ in the large $n$ regime
and the CLT. Due to the Gaussian character of the entries of Matrix ${\bf
  G}$, two simple ingredients are available. The first one is an
Integration by parts formula (\ref{eq-integ-parts}) that provides an
expression for the expectation of certain functionals of Gaussian
vectors. This formula has been widely used in RMT 
\cite{KhoPas93,KPV95,pas-umj05}. The second
ingredient is Poincar\'e-Nash inequality (\ref{eq-np}) that bounds the
variance of functionals of Gaussian vectors. Although well known 
\cite{Che82,HPS98},
its application to RMT is fairly recent \cite{pas-umj05}. This
inequality enables us to control the decrease rate of
the approximation errors
such as the order $1/n$ error $\EE\left[ I(\rho) \right] - V(\rho)$. 
We believe that these tools which prove to be simple and robust 
might be of great interest for the communications engineering community. 

The paper is organized as follows. In Section \ref{sec-problem}, we
introduce the main notations; we also state the two main results of
the article. In Section \ref{sec-tools}, we recall general matrix
results and the two aforementioned Gaussian tools. Section
\ref{sec-first-order} is devoted to the proof of the first order
result, that is the approximation of $\EE[I(\rho)]$. The CLT, also
refered to as the second order result, is established in Section
\ref{sec-second-order-clt}. Proof details are in an appendix.

\section{Notations and Statement of the main results}
\label{sec-problem}

\subsection{From a Kronecker model to a separable variance model.}
Consider a MIMO system represented by a $N \times n$ matrix ${\bf G}$ where
$n$ is the number of antennas at the transmitter and $N$ is the number
of antennas at the receiver and where $N(n)$ is a sequence of integers such 
that 
$$
\lim_{n\to\infty} \frac{N(n)}{n} = c > 0. 
$$
Assuming the transmitted signal is a Gaussian signal with a covariance
matrix equal to $\frac{1}{n} {\bf I}_n$ (and thus, a total power equal
to one), Shannon's mutual information of this channel is $ I_n(\rho) =
\log\det\left( \frac{\rho}{n} {\bf GG}^* + {\bf I}_N \right), $ where
$\rho>0$ is the inverse of the additive white Gaussian noise variance
at each receive antenna.  The general problem we address in this paper
concerns the behaviour of the mutual information for large values of
$N$ and $n$ in the case where the channel matrix ${\bf G}$, assumed to
be random, is described by the Kronecker model ${\bf G} = {\bs \Psi}
{\bf W} \widetilde{\bs \Psi}$. In this model, ${\bs \Psi}$ and
$\widetilde{\bs \Psi}$ are respectively $N \times N$ and $n \times n$
deterministic matrices and ${\bf W}$ is random with independent
entries distributed acccording to the complex circular Gaussian law
with mean zero and variance one ${\cal CN}(0,1)$.

It is well known that this model can be replaced by a simpler
Kronecker model involving a matrix with Gaussian independent (but not
necessarily identically distributed) entries.  Indeed, let ${\bs \Psi}
= {\bf U} {\bf D}_n^{\frac 12} {\bf V}^*$ (resp. $\widetilde{\bs \Psi} =
\widetilde{\bf U} \widetilde{\bf D}_n^{\frac 12} \widetilde{\bf V}^*$) be a
Singular Value Decomposition (SVD) of ${\bs \Psi}$ (resp.
$\widetilde{\bs \Psi}$), where ${\bf D}_n$ (resp. $\widetilde{\bf D}_n$)
is the diagonal matrix of eigenvalues of ${\bs \Psi} {\bs \Psi}^*$
(resp. $\widetilde{\bs \Psi} \widetilde{\bs \Psi}^*$), then $I_n(\rho)$ writes:
$$
I_n(\rho) = \log\det\left( \frac{\rho}{n} {\bf Y}_n {\bf Y}_n^* + {\bf I}_N
\right),
$$
where ${\bf Y}_n = {\bf D}_n^{\frac 12} {\bf X}_n \widetilde{\bf D}_n^{\frac 12}$ is a
$N\times n$ matrix, ${\bf D}_n$ and $\widetilde {\bf D}_n$ are respectively $N \times
N$ and $n \times n$ diagonal matrices, i.e.
$$
{\bf D}_n=\mathrm{diag} \left(d_i^{(n)},\ 1\le i\le N\right) \quad \textrm{and} \quad 
\widetilde{\bf D}_n=\mathrm{diag} \left(\tilde d_j^{(n)},\ 1\le j\le n\right),
$$ 
and ${\bf X}_n={\bf V}^* {\bf W} \widetilde{\bf U}$ has i.i.d. entries
with distribution ${\cal CN}(0,1)$ since ${\bf V}$ and $\widetilde{\bf
  U}$ are deterministic unitary matrices. Since every individual entry
of ${\bf Y}_n$ has the form $Y_{ij}^{(n)} = \sqrt{d_i^{(n)}
  \tilde{d}_j^{(n)}} X_{ij}$, we call ${\bf Y}_n$ a random matrix with a
separable variance profile.

\subsection{Assumptions and Notations.}

The centered random variable $X - \EE[X]$ will be denoted by
$\overcirc{X}$. Element $(i,j)$ of a matrix ${\bf A}$ will be either
denoted $[ {\bf A} ]_{ij}$ or $A_{ij}$.  Element $i$ of vector ${\bf
  a}$ will be denoted $a_i$ or $[ {\bf a} ]_i$.  Column $j$ of matrix
${\bf A}$ will be denoted ${\bf a}_j$.  The transpose, the Hermitian
adjoint (conjugate transpose) of ${\bf A}$, and the matrix obtained by
conjugating its elements are denoted respectively ${\bf A}^T$, ${\bf
  A}^*$, and $\overline{\bf A}$.  The spectral norm of a matrix ${\bf
  A}$ will be denoted $\| {\bf A} \|$.  If ${\bf A}$ is square, $\tr
{\bf A}$ refers to its trace.  Let $\mathbf{i}=\sqrt{-1}$, then the
operators $\partial / \partial z$ and $\partial / \partial
\overline{z}$ where $z = x+ \mathbf{i} y$ is a complex number are
defined by $\frac{\partial}{\partial z} = \frac 12 \left(
  \frac{\partial}{\partial x} - \mathbf{i} \frac{\partial}{\partial y}
\right)$ and $\frac{\partial}{\partial \overline{z}} = \frac 12 \left(
  \frac{\partial}{\partial x} + \mathbf{i} \frac{\partial}{\partial y}
\right)$
where $\frac{\partial}{\partial x}$ and $\frac{\partial}{\partial y}$ 
are the standard partial derivatives with respect to $x$ and $y$. \\

Throughout the paper, notation $K$ will denote a generic constant
whose main feature is {\em not} to depend on $n$. In particular, the
value of $K$ might change from a line to another as long as it never
depends upon $n$.  Constant $K$ might depend on $t\in \mathbb{R}^+$
and whenever needed, this dependence will be made more explicit. \\ 
As usual notation $\alpha_n={\mathcal O}(\beta_n)$ is a flexible
shortcut for $|\alpha_n|\le K \beta_n$ and $\alpha_n=
  o(\beta_n)$, for $\alpha_n =\varepsilon_n \beta_n$ with
$\varepsilon_n\rightarrow 0$ as $n$ goes to infinity.

In order to study a deterministic approximation of $I_n(\rho)$ and its
fluctuations, the following mild assumptions are required over the two
triangular arrays $\left( d_i^{(n)},\ 1\le i\le N,\ n\ge 1\right)$
and $\left( \tilde d_j^{(n)}, 1\le j\le n,\ n\ge 1 \right)$.
\begin{description}
\item[{\bf (A1)}] The real numbers $d_i^{(n)}$ and $\tilde d_j^{(n)}$ are
      nonnegative and the sequences $\left( d_i^{(n)} \right)$ and
      $\left( \tilde d_j^{(n)} \right)$ are uniformly bounded, i.e.
      there exist constants $d_{\mathrm{max}}$ and
      $\tilde d_{\mathrm{max}}$ such that 
$$ 
\sup_n \| {\bf D}_n \|
      < d_{\mathrm{max}} \quad \mathrm{and} \quad \sup_n \|
      \widetilde{\bf D}_n \| < \tilde d_{\mathrm{max}}.
$$
where $\| {\bf D}_n \|$ and $\|\widetilde{\bf D}_n \|$
are the spectral norms of ${\bf D}_n$ and $\widetilde{\bf D}_n$.
% \sup_n \max_{i=1, \ldots, N} d_i^{(n)} < d_{\mathrm{max}} \quad 
% \mathrm{and} \quad 
% \sup_n \max_{j=1, \ldots, n} \tilde d_j^{(n)} < \tilde d_{\mathrm{max}}
\item[{\bf (A2)}] The normalized traces of ${\bf D}_n$ and 
$\widetilde{\bf D}_n$ satisfy 
$$
\inf_n \frac{1}{n} \tr \left( {\bf D}_n \right) > 0 
\quad \mathrm{and} \quad 
\inf_n \frac{1}{n} \tr \left( \widetilde{\bf D}_n \right) > 0. 
$$
\end{description} 
In the sequel, we shall frequently omit the subscript $n$ and the 
superscript $(n)$. \\ 
The resolvent associated with $ \frac{1}{n} {\bf Y}_n {\bf Y}_n^*$ is
the $N \times N$ matrix ${\bf H}_n(t) = \left( \frac tn {\bf Y}_n 
{\bf Y}_n^* + {\bf I}_N \right)^{-1}$. Of prime importance is the random 
variable 
$\beta(t) = \frac{1}{n} \tr {\bf DH}(t) $ and its expectation $\alpha(t) = \frac{1}{n} 
\tr {\bf D} \, \EE {\bf H}(t)$. \\
We furthermore introduce the $n\times n$ deterministic matrix defined by
\begin{eqnarray*}
\widetilde{\bf R}(t) &=& \left({\bf  I} +t\alpha(t) \widetilde{\bf D}_n\right)^{-1}, \\
&=& \diag\left( \tilde{r}_j(t),\ 1\le j\le n\right)\quad \textrm{where}\quad 
\tilde{r}_{j}(t) = \frac{1}{ 1 + t \alpha(t) \tilde{d}_j },
\end{eqnarray*}
and the related quantity $\tilde{\alpha}(t) = \frac 1n \tr 
\widetilde{\bf D} \widetilde{\bf R}(t)$.
In a symmetric fashion, the $N\times N$ matrix ${\bf R}(t)$ is defined by
\begin{eqnarray*}
{\bf R}(t) &=& \left({\bf  I} +t\tilde\alpha(t) {\bf D}_n\right)^{-1},\\
&=& \diag\left( r_i(t),\ 1\le i\le N\right)\quad \textrm{where}\quad 
r_{i}(t) = \frac{1}{ 1 + t \tilde \alpha(t) d_i }.
\end{eqnarray*}
We finally introduce the solutions of a deterministic $2\times 2$ system.

\begin{proposition}
\label{existence-unicite}
For every $n$, the system of equations in $(\delta, \tilde\delta)$ 
\begin{equation}
\label{eq-equations-canoniques}
\left\{ 
\begin{array}{ccc}
\delta & = & 
\frac{1}{n} \tr 
{\bf D}_n ({\bf I} + t \tilde{\delta} {\bf D}_n)^{-1} \\
\tilde{\delta} & = & 
\frac{1}{n} \tr 
\widetilde{{\bf D}}_n ({\bf I} + t \delta \widetilde{{\bf D}}_n)^{-1} 
\end{array}\right. 
\end{equation}
admits a unique solution $\left( \delta_n(t), \tilde\delta_n(t) \right)$ 
satisfying $\delta_n(t) > 0, \tilde\delta_n(t) > 0$. Moreover, there exist
nonnegative measures $\mu_n$ and $\tilde\mu_n$ over $\mathbb{R}^+$ such that 
\begin{equation}\label{representation-stieltjes}
\delta_n(t)=\int_{\mathbb{R}^+}
\frac{\mu_n(d\lambda)}{1+t\lambda}\qquad \textrm{and}\qquad
\tilde\delta_n(t)=\int_{\mathbb{R}^+}
\frac{\tilde\mu_n(d\lambda)}{1+t\lambda}\ ,
\end{equation}

where $\mu_n(\mathbb{R}^+)=\frac 1n \tr {\bf D}_n$ and 
$\tilde\mu_n(\mathbb{R}^+)=\frac 1n \tr \widetilde{\bf D}_n$.
\end{proposition}

The proof is postponed to Appendix \ref{proof-existence-unicite}.

With $\delta$ and $\tilde \delta$ properly defined, we introduce the 
following $N\times N$ and $n\times n$ diagonal matrices:
$$
{\bf T}=({\bf I} +t\tilde\delta {\bf D})^{-1} \quad \textrm{and}\quad 
\widetilde{\bf T}=({\bf I} +t\delta \widetilde{\bf D})^{-1}.
$$ 
Notice in particular that $\delta=\frac 1n \tr\, {\bf D}{\bf T}$ and 
$\tilde \delta =\frac 1n \tr\, \widetilde{\bf D} \widetilde{\bf T}$ by \eqref{eq-equations-canoniques}.   
We finally introduce the following quantities which are required to 
express the fluctuations of $I_n(\rho)$:
\begin{equation}\label{eq-var-var}
\left\{
\begin{array}{l}
\gamma_n(t) = \frac{1}{n} 
\tr {\bf D}^2_n {\bf T}^2_n(t) \phantom{\bigg(} \\ 
\tilde{\gamma}_n(t)  =  
\frac{1}{n} 
 \tr  \widetilde{{\bf D}}^2_n \widetilde{{\bf T}}^2_n(t) 
\end{array}\right. \ .
\end{equation}

\begin{proposition}\label{existence-variance}
Assume that Assumptions {\bf (A1)} and {\bf (A2)} hold and denote by 
\begin{equation}\label{definition-variance}
\sigma_n^2\left(t\right) = - \log\left( 1 - t^2 \gamma_n(t) \tilde\gamma_n(t) \right),\quad t>0
\end{equation}
where $\gamma_n(t)$ and $\tilde\gamma_n(t)$ are given by \eqref{eq-var-var}.
Then $\sigma^2_n(t)$ is well-defined, i.e. $1 - t^2 \gamma_n(t) \tilde\gamma_n(t)>0$ for $t>0$. 
Moreover there exist nonnegative real numbers $m_{t}$ and $M_{t}$ such that
\begin{equation}\label{controle-variance}
0<m^2_{t}\le \inf_n \sigma^2_n(t) \le \sup_n \sigma_n^2(t) \le M^2_{t}< \infty\quad \textrm{for}\quad t>0\ .
\end{equation}
Moreover, $\sigma_n^2(t)$ is upper-bounded uniformly in $n$ and $t$ for $t\in [0,\rho]$, i.e. 
$\sup_{t\le \rho} M^2_t <\infty$.\\
\end{proposition}

Proof of Proposition \ref{existence-variance} is postponed to Appendix \ref{preuve-existence-variance}. 

\subsection*{Summary of the main notations.}

In order to improve the readability of the paper, we gather all the
notations in Table \ref{resume-notations}. As expressed there, there
are three kinds of quantities: 
\begin{enumerate}
\item Random quantities, 
\item Deterministic
quantities depending on the law of ${\bf Y Y^*}$ via the expectation $\EE$
with respect to the entries of ${\bf Y}$, 
\item Deterministic quantities which only depend on the matrices 
${\bf D}$ and $\widetilde{\bf D}$, sometimes
via $\delta$ and $\tilde\delta$ (as defined in Proposition
\ref{existence-unicite}) which are easily computable.
\end{enumerate}
The main goal of the forthcoming computations will be to approximate
elements of the first and second kind by elements of the third kind.

\begin{table}\label{resume-notations}
\begin{center}
\newlength{\LLL}\settowidth{\LLL}{Random quantities}
\setlength{\extrarowheight}{4pt}
\begin{tabular}{|l|l|l|}
  \hline
  \multirow{2}{\LLL}{Random quantities} 
  &\multicolumn{2}{c|}{Deterministic quantities}\\\cline{2-3}
  & depending on the law of ${\bf YY^*}$ via $\EE$ & only depending on the variance structure via ${\bf D}$ and $\widetilde{\bf D}$\\
  \hline \hline
  ${\bf H}= \left( \frac tn {\bf Y Y^*} +{\bf I}\right)^{-1}$ &&\\
  $\beta =\frac 1n \tr {\bf D} {\bf H}$ & $\alpha=  \frac 1n \tr {\bf D} (\EE {\bf H})$
  & $\delta =\frac 1n \tr {\bf D} ({\bf I} +t\tilde\delta {\bf D})^{-1}=\frac 1n \tr {\bf D} {\bf T}$\\ 
  & $\tilde r_j =(1+t\alpha \tilde d_j)^{-1}$ &\\
  & $\widetilde{\bf R} =({\bf I} +t\alpha\widetilde{\bf D})^{-1}$ 
  &$\widetilde{\bf T} =({\bf I} +t\delta\widetilde{\bf D})^{-1}$\\
  &$\tilde \alpha =\frac 1n \tr \widetilde{\bf D} \widetilde{\bf R}
  =\frac 1n \tr \widetilde{\bf D}({\bf I} +t\alpha\widetilde{\bf D})^{-1} $ & 
  $\tilde\delta  
  =\frac 1n \tr \widetilde{\bf D}({\bf I} +t\delta\widetilde{\bf D})^{-1}=\frac 1n 
  \tr \widetilde{\bf D} \widetilde{\bf T}$   \\
  & $r_i = (1+t \tilde \alpha d_i)^{-1}$ &\\
  & ${\bf R} = ({\bf I} +t\tilde \alpha {\bf D})^{-1}$ & ${\bf T} = ({\bf I} +t\tilde \delta {\bf D})^{-1}$\\
  && $\gamma =\frac 1n \tr {\bf T}^2{\bf D}^2,\quad $ 
$\tilde{\gamma}=\frac 1n \tr \widetilde{\bf T}^2 \widetilde{\bf D}^2$\\   
  && $\sigma^2(t) =-\log (1-t^2 \gamma(t) \tilde \gamma(t))$\\
  \hline
\end{tabular}
\end{center}
\caption{summary of the main notations}
\end{table}

\subsection{Statement of the main results.}
We now state the main results. Theorem \ref{theo-ordre-1} describes the first order 
approximation of the Shannon capacity $I_n(\rho)$ while Theorem \ref{theo-clt} describes its 
fluctuations when centered with respect to its first order approximation.
\begin{theorem}
\label{theo-ordre-1}
Let ${\bf X}$ be a $N\times n$ matrix whose elements $X_{ij}$ are
independent complex Gaussian variables such that
$$
\EE (X_{ij})= \EE(X_{ij}^2)=0,\quad \EE(|X_{ij}|^2)=1,\quad 1\le i\le N,\ 1\le j\le n,
$$
and ${\bf Y}={\bf D}^{\frac 12} {\bf X} \widetilde{\bf D}^{\frac 12}$
where the diagonal matrices ${\bf D}$ and $\widetilde{\bf D}$ satisfy
Assumptions {\bf (A1)} and {\bf (A2)}. Let $I_n(\rho) = \log \det
\left( \frac {\rho}{n} {\bf Y} {\bf Y}^* + {\bf I}_N \right)$.
Then, we have 
\begin{equation}
\label{eq-expreI}
\EE [ I_n(\rho) ] = V_n(\rho) + {\cal O}\left(\frac{1}{n}\right) 
\end{equation}
as $n\rightarrow \infty$, $Nn^{-1} \rightarrow c\in ]0,\infty[$ where 
$$
V_n(\rho) =  
\log \det \left( {\bf I} + \rho \delta_n(\rho) \widetilde{{\bf D}}_n \right) 
+ \log \det \left( {\bf I} + \rho \tilde{\delta}_n(\rho) {\bf D}_n \right) 
- n \rho \delta_n(\rho) \tilde{\delta}_n(\rho) \ . 
$$
and where $(\delta_n(t),\tilde{\delta}_n(t))$ is the unique positive solution of the system
$$
\left\{ 
\begin{array}{ccc}
\delta & = & 
\frac{1}{n} \tr 
{\bf D} ({\bf I} + t \tilde{\delta} {\bf D})^{-1} \\
\tilde{\delta} & = & 
\frac{1}{n} \tr 
\widetilde{{\bf D}} ({\bf I} + t \delta \widetilde{{\bf D}}_n)^{-1} 
\end{array}\right. .
$$

\end{theorem}
\vspace{0.03\columnwidth} 

\begin{theorem}
\label{theo-clt} 
Assume that the setting of Theorem \ref{theo-ordre-1} holds and let
$
\sigma^2_n(\rho)=-\log\left( 1 - \rho^2 \gamma_n(\rho) \tilde\gamma_n(\rho) \right)\ .
$
Then the random variable 
$\sigma_n^{-1}(\rho) (I_n(\rho) - V_n(\rho))$ 
converges in distribution towards ${\cal N}(0,1)$ where
$$
\left\{
\begin{array}{l}
\gamma_n(\rho) = \frac{1}{n} 
\tr {\bf D}^2_n {\bf T}^2_n(\rho)\\
\tilde{\gamma}_n(\rho)  =  
\frac{1}{n} 
\tr  \widetilde{{\bf D}}^2_n \widetilde{{\bf T}}^2_n(\rho) 
\end{array}\right. 
\quad \textrm{and}\
\left\{
\begin{array}{l}
{\bf T}(\rho)=({\bf I} +\rho \tilde \delta {\bf D})^{-1}\\
\widetilde{\bf T}(\rho)=({\bf I} +\rho \delta \widetilde{\bf D})^{-1}
\end{array}\right. .
$$

\end{theorem}
\section{Mathematical Tools and Some Useful Results} 
\label{sec-tools}

In this section, we present the tools we will use extensively all
along the paper. In Section \ref{sec-generalites}, we recall
well known matrix results; in Section \ref{sec-gaussian-tools}, we
present two fundamental properties of Gaussian models: The Integration
by parts formula and Poincar\'e-Nash inequality for Gaussian vectors.
Section \ref{sec-cornerstone} is devoted to a cornerstone
approximation result which roughly states that ${\bf R}$ and
$\widetilde{\bf R}$ can be replaced by ${\bf T}$ and $\widetilde{\bf
  T}$ up to some well-quantified error. In Section \ref{more-approx},
various variance estimates and approximation rules are stated.

\subsection{General results}\label{sec-generalites}

\subsubsection{Some matrix inequalities}

Let ${\bf A}$ and ${\bf B}$ be two $N \times N$ matrices with complex 
elements. Then 
% \begin{equation}
% \label{eq-(trA)^2<Ntr(A^2)} 
% \left| \tr\left( {\bf A} \right) \right|^2 \leq 
% N \tr\left( {\bf AA}^* \right) \ . 
% \end{equation} 
\begin{equation}
\label{eq-tr(ab)<tr(aa)tr(bb)} 
\left| \tr\left({\bf AB}\right) \right| \leq 
\sqrt{\tr\left( {\bf AA}^* \right)}
\sqrt{\tr\left( {\bf BB}^* \right)}  \ . 
\end{equation} 
Assuming ${\bf A}$ is Hermitian nonnegative, we have 
\begin{equation}
\label{eq-tr(ab)<|b|tr(a)} 
\left| \tr \left( {\bf A} {\bf B} \right) \right| \leq 
\| {\bf B} \| \ \tr \left( {\bf A} \right)\ , 
\end{equation} 
where $\| . \|$ is the spectral norm (see \cite{hor-joh-livre94}). \\

\subsubsection{The Resolvent}
The Resolvent matrix ${\bf H}_n(t)$ of matrix ${\bf Y}_n {\bf Y}_n^*$
is defined as ${\bf H}_n(t) =\left( \frac tn {\bf Y}_n {\bf Y}_n^* +
  {\bf I}_N \right)^{-1}$.  It is of constant use in this paper and we
give here some of its properties. The following identity, also known
as the {\em Resolvent identity}:
\begin{equation}
\label{eq-resolvent-id} 
{\bf H}(t) = {\bf I}_N - \frac{t}{n} {\bf H}(t) {\bf YY}^* 
\end{equation} 
follows from the mere definition of ${\bf H}$.
Furthermore, the spectral norm of the 
resolvent is readily bounded by one: 
\begin{equation}
\label{eq-|H|<1} 
\| {\bf H}(t) \| \leq 1 \quad \mathrm{for} \quad t \geq 0 \ . 
\end{equation} 

\subsubsection{Bounded character of the mean of some empirical moments} 

Let $({\bf B}_n)_{n\in \mathbb{N}} = \diag\left(\left[ b_1^{(n)},
    \ldots, b_n^{(n)} \right] \right)$, $n\in \mathbb{N}$, be a sequence
of deterministic $n \times n$ diagonal matrices.  Assume {\bf (A1)},
and furthermore, that $\sup_n \| {\bf B}_n \| < \infty$.  Then for
every integer $k$, we have
\begin{equation}
\label{eq-trace-bounded} 
\frac 1n \EE \left[ \tr 
\left( \frac 1n {\bf Y} {\bf B} {\bf Y}^* \right)^k 
\right] < K \ .
\end{equation} 
Let us sketch a proof. Expanding the left hand side of (\ref{eq-trace-bounded}) yields:
$$
\frac{1}{n^{k+1}} 
\sum_{\genfrac{}{}{0pt}{}{i_1, i_2, \ldots, i_k = 1:N}
{j_1, \ldots, j_k = 1:n}}
b_{j_1} b_{j_2} \cdots b_{j_k} 
\EE\left[
Y_{i_1 j_1} \overline{Y_{i_2 j_1}} 
Y_{i_2 j_2} \overline{Y_{i_3 j_2}} 
\cdots 
Y_{i_k j_k} \overline{Y_{i_1 j_k}} 
\right]\ .
$$
A close look at the argument of the $\EE$ operator implies that due to 
the independence of the $Y_{ij}$, we only
have $k+1$ degrees of freedom in the choice of the indices 
$i_p$ and $j_q$. As all moments of the Gaussian law exist and moreover 
$\| {\bf B}_n \|$, $\| {\bf D}_n \|$, and $\| \widetilde{\bf D}_n \|$ are
bounded, this sum is of order $1$ as $n \to \infty$. \\

\subsubsection{Differentiation formulas} 

Let ${\bf A}$ be a $N \times N$ complex matrix and let  
${\bf Q}({\bf A}) = \left( {\bf I}_N + {\bf A} \right)^{-1}$. 
Let ${\bs \delta}{\bf A}$ be a perturbation of ${\bf A}$. Then
\begin{equation} 
\label{eq-differential-resolvent}
{\bf Q}({\bf A} + {\bs \delta}{\bf A} ) = {\bf Q}({\bf A}) 
- {\bf Q}({\bf A}) \ {\bs \delta}{\bf A} \ {\bf Q}({\bf A}) + 
o\left( \| {\bs \delta}{\bf A} \| \right), 
\end{equation} 
where $o\left( \| {\bs \delta}{\bf A} \| \right)$ is negligible with respect
to $\| {\bs \delta}{\bf A} \|$ in a neighborhood of $0$.
% \begin{equation} 
% \label{eq-differential-resolvent}
% {\bf G}'({\bf A}) \ {\bf .} \ {\bf X} 
% = - {\bf G}({\bf A}) {\bf X} {\bf G}({\bf A}) \ .
% \end{equation} 
Writing ${\bf H}(t) = \left[ H_{pq}(t) \right]_{p,q=1}^{N,N}$, we need
the expression of the partial derivative $\partial H_{pq} / \partial Y_{ij}$. 
Using (\ref{eq-differential-resolvent}), we have:  
\begin{eqnarray}
\frac{\partial H_{pq}}{\partial Y_{ij}} & = & 
- \frac{t}{n} 
\left[
{\bf H} \frac{\partial {\bf YY}^*}{\partial Y_{ij}} {\bf H} 
\right]_{pq} 
=
- \frac{t}{n} 
\left[
{\bf H}  
\left[ \phantom{\widetilde X^2} \! \! \! \! \! \! \!   
 \delta(k-i) \overline{Y_{\ell j}} \right]_{k,\ell=1}^N 
{\bf H} 
\right]_{pq} \nonumber \\
&=& 
- \frac tn H_{pi} \left[ {\bf Y}^* {\bf H} \right]_{jq} 
= -\frac tn H_{pi} [{\bf y}_j^* {\bf H}]_q \ ,
\label{eq-dH/dY} 
\end{eqnarray}
where $\delta$ is the Kronecker function. 
Similarly, we can establish 
\begin{equation}
\label{eq-dH/dY^*}
\frac{\partial H_{pq}}{\partial \overline{Y_{ij}}} = 
- \frac tn \left[ {\bf H} {\bf Y} \right]_{pj} H_{iq} 
= -\frac tn [{\bf H}{\bf y}_j]_p H_{iq} 
\ . 
\end{equation} 

The differential of $g({\bf A}) = \log\det({\bf A})$ is given by
$
% g'({\bf A}) \ {\bf .} \ {\bf X} = \tr\left( {\bf A}^{-1} {\bf X} \right) \ .
g({\bf A} + {\bs \delta}{\bf A}) = 
g({\bf A}) + \tr\left( {\bf A}^{-1} \ {\bs \delta}{\bf A} \right) 
+ o\left( \| {\bs \delta}{\bf A} \| \right) \ . 
$
We use this equation to derive the expression of 
$\partial I(t) / \partial \overline{Y_{ij}}$ also needed below: 
\begin{equation}
\label{eq-dI/dY^*} 
\frac{\partial I}{\partial \overline{Y_{ij}}} = 
\frac tn \tr\left(
{\bf H} \frac{\partial {\bf YY}^*}{\partial \overline{Y_{ij}}} \right)
= 
\frac tn \tr\left( {\bf H} 
\left[ \phantom{\widetilde X^2} \! \! \! \! \! \! \!   
\delta(\ell-j) Y_{kj} \right]_{k,\ell=1}^N \right) 
= 
\frac{t}{n} \left[ {\bf H Y} \right]_{ij}
= \frac tn \left[ {\bf Hy}_j \right]_i \ .
\end{equation}

\subsection{Gaussian tools}\label{sec-gaussian-tools}

\subsubsection{An Integration by parts formula for Gaussian functionals}
Let ${\bs \xi} = [ \xi_1, \ldots, \xi_M ]^T$ be a complex Gaussian
random vector whose law is determined by $\EE [ {\bs \xi} ] = {\bf
  0}$, $\EE [ {\bs \xi} {\bs \xi}^T ] = {\bf 0}$, and $\EE [ {\bs \xi}
{\bs \xi}^* ] = {\bs \Xi}$. Let $\Gamma=\Gamma(\xi_1,\cdots,\xi_M,\overline{\xi_1},\cdots, 
\overline{\xi_M})$ 
be a ${\mathcal  C}^1$ complex function polynomially bounded together with 
its derivatives, then:
\begin{equation}
\label{eq-integ-parts}
\EE \left[ \xi_p \Gamma({\bs \xi}) \right] = 
\sum_{m=1}^M \left[ {\bs \Xi} \right]_{pm} 
\EE \left[ \frac{\partial \Gamma({\bs \xi})}{\partial \overline{\xi_{m}}} \right]\ .
\end{equation} 
This formula relies on an integration by parts and thus is 
referred to as
the Integration by parts formula for Gaussian vectors. It is widely used in Mathematical Physics
\cite{GliJaf87} and has been used in Random Matrix Theory in \cite{KhoPas93,KPV95}.\\

\subsubsection{Poincar\'e-Nash inequality}
Let ${\bs \xi}$ and $ \Gamma$ be as previously and let 
 $\nabla_z\Gamma = [ \partial\Gamma / \partial z_1, \ldots, 
 \partial\Gamma / \partial z_M ]^T$ and 
 $\nabla_{\overline{z}}\Gamma = [ \partial\Gamma / 
 \partial \overline{z_1}, \ldots, 
 \partial\Gamma / \partial \overline{z_M} ]^T$. 
Then the following inequality holds true:
\begin{equation}
\label{eq-np} 
 \mathrm{var}\left( {\Gamma}({\bs \xi}) \right) \leq 
 \EE \left[ \nabla_z \Gamma({\bs \xi})^T \ {\bs \Xi} \ 
 \overline{\nabla_z \Gamma({\bs \xi})} 
 \right] 
 +
 \EE \left[ \left( \nabla_{\overline{z}}  \Gamma({\bs \xi}) \right)^* 
 \ {\bs \Xi} \ 
 \nabla_{\overline{z}} \Gamma({\bs \xi}) 
 \right] 
 \ .
\end{equation}
This inequality is well-known (see e.g. \cite{Che82,HPS98}) and has 
first been applied to Random Matrix Theory in \cite{pas-umj05}.

When ${\bs \xi}$ is the vector of the stacked columns of matrix ${\bf Y}$, i.e.
${\bs \xi} = [ Y_{11}, \ldots, Y_{Nn} ]^T$, formula \eqref{eq-integ-parts} becomes:
\begin{equation}
\label{eq-integ-parts-our-model}
\EE \left[ Y_{ij}  \Gamma({\bf Y}) \right] = 
d_i \tilde{d}_j  
\EE \left[ \frac{\partial \Gamma({\bf Y})}{\partial \overline{Y_{ij}}} 
\right] \ , 
\end{equation} 
while inequality \eqref{eq-np} writes:
\begin{equation}
\label{eq-np-our-model} 
\mathrm{var}\left(\Gamma({\bf Y})\right) \leq 
\sum_{i=1}^N \sum_{j=1}^n d_i \tilde d_j  
\EE \left[ 
\left| \frac{\partial \Gamma({\bf Y})}
{\partial Y_{i,j}} \right|^2 
+ 
\left| \frac{\partial \Gamma({\bf Y})}
{\partial \overline{Y_{i,j}}} \right|^2 
\right] \ .
\end{equation}

Poincar\'e-Nash inequality turns out to be extremely useful to deal
with variances of various quantities of interest related with
random matrices. For the reader's convenience, we provide a proof in
Appendix \ref{appendix-poincare} and in order to give right away the
flavour of such results, we state and prove the following:

\begin{proposition}\label{PN-controle-1} Assume that the setting of Theorem 
\ref{theo-ordre-1} holds and let ${\bf A}_n$ be a $N\times N$ real
diagonal matrix which spectral norm is uniformly bounded in $n$. Then 
$$
\mathrm{var}\left( \frac 1n\tr {\bf A}{\bf H} \right) = 
{\cal O}\left(n^{-2} \right)\ .
$$
\end{proposition}
\begin{proof}
We apply inequality (\ref{eq-np-our-model}) to 
the function $\Gamma({\bf Y})=\frac 1n \tr {\bf AH}$.
% given by the statement of the Lemma. 
Using (\ref{eq-dH/dY}), we have 
$$ 
\frac{\partial \Gamma}{\partial Y_{i,j}}  = 
\frac 1n \sum_{p=1}^N a_p \frac{\partial H_{pp}}{\partial Y_{i,j}}  
=
- \frac{t}{n^2} [ {\bf y}_j^* {\bf HAH} ]_{i} \ . 
$$ 
Therefore, denoting by $A$ the upper bound $A = \sup_n \| {\bf A}_n \|$ and
noticing that $\left| \partial \Gamma / \partial Y_{i,j} \right| = 
\left| \partial \Gamma / \partial \overline{Y_{i,j}} \right|$,  
we have:
\begin{eqnarray*} 
\mathrm{var}\, \Gamma({\bf Y})  &\leq& 
\frac{2 t^2}{n^4}
\sum_{i=1}^N \sum_{j=1}^n d_i \tilde d_j \EE \left| \left[ {\bs y}_j^* {\bf HAH}\right]_i\right|^2 \\
&=& 
\frac{2 t^2}{n^4} \sum_{j=1}^n \tilde{d}_j 
\EE\left( {\bf y}_j^* {\bf HAHDHAHy}_j  \right)  \\ 
&=& 
\frac{2 t^2}{n^3} \EE  
\,\tr \left(
{\bf HAHDHAH}  \frac{{\bf Y}\widetilde{\bf D} {\bf Y}^*}{n}
\right)  \\
&\stackrel{(a)}{\leq}& 
\frac{2 t^2}{n^3} \EE \left\{ 
\| {\bf H} \|^4 \| {\bf A} \|^2 \| {\bf D} \| \ 
\tr \left( \frac{{\bf Y}\widetilde{\bf D} {\bf Y}^*}{n}
\right) \right\} \quad \stackrel{(b)}{\leq}\quad
\frac{2 A^2 d_{\mathrm{max}} t^2}{n^3} \EE \,
\tr \left( \frac{{\bf Y}\widetilde{\bf D} {\bf Y}^*}{n}
\right) 
\quad \stackrel{(c)}{\le}\quad \frac{K}{n^2} \ ,
\end{eqnarray*} 
where inequality $(a)$ follows from (\ref{eq-tr(ab)<|b|tr(a)}), 
$(b)$ follows from (\ref{eq-|H|<1}) and from the bounded character of 
$\| {\bf A}_n \|$ and $\| {\bf D}_n \|$, and $(c)$ follows from
(\ref{eq-trace-bounded}).  
\end{proof}

%\subsubsection{The main estimates}
%\begin{lemma} Let $({\bf A}_n)$ and $({\bf B}_n)$ be two sequences of respectively 
%$N\times N$ and $n\times n$ diagonal deterministic matrices whose spectral norm 
%are uniformly bounded in $n$, then the following hold true:
%\begin{eqnarray}
%&\mathrm{var}\left( \frac 1n \tr {\bf A}{\bf H} \right) &= {\cal O}\left(\frac 1{n^2}\right)\\
%&\mathrm{var}\left( \frac 1n \tr {\bf A}{\bf H}\left( \frac{{\bf Y} {\bf B}{\bf Y^*}}{n}\right) \right)
%&= {\cal O}\left(\frac 1{n^2}\right)\\
%&\mathrm{var}\left( \frac 1n \tr {\bf A}{\bf H} {\bf D} {\bf H}\left( \frac{{\bf Y} {\bf B}{\bf Y^*}}{n}\right) \right)
%&= {\cal O}\left(\frac 1{n^2}\right)
%\end{eqnarray}

%\end{lemma}

\subsection{Approximation rules}\label{sec-cornerstone}
The following theorem is crucial in order to prove Theorems \ref{theo-ordre-1} and \ref{theo-clt}. Roughly speaking 
it allows to replace matrices ${\bf R}$ and $\widetilde{\bf R}$ by ${\bf T}$ and $\widetilde{\bf T}$ up
to a well-quantified small error.
\begin{theorem}\label{theo-canonique}
Let $({\bf A}_n)$ and $({\bf B}_n)$ be two sequences of respectively 
$N\times N$ and $n\times n$ diagonal deterministic matrices whose spectral norm 
are uniformly bounded in $n$, then the following hold true:
\begin{eqnarray}
  \frac 1n \tr {\bf A} {\bf R}  & =& \frac 1n \tr {\bf A} {\bf T} + {\cal O}\left(\frac 1{n^2}\right),\label{approx-uno}\\
  \frac 1n \tr {\bf B} \widetilde{\bf R} & =& \frac 1n \tr {\bf B} \widetilde{\bf T} 
+ {\cal O}\left(\frac 1{n^2}\right).
\end{eqnarray}

\end{theorem}
Proof of Theorem \ref{theo-canonique} is postponed to Appendix 
\ref{anx-proof-theo-canonique}.

\subsection{More variance estimates and more approximations rules}\label{more-approx}

We collect here a few results which proofs rely on the Integration by
parts formula \eqref{eq-integ-parts-our-model}, on Poincar\'e-Nash
inequality, and on Theorem \ref{theo-canonique}. The proofs of these
results, although systematic, are somewhat lengthy and are therefore postponed 
to the Appendix. These results will be used extensively in Section
\ref{sec-second-order-clt}. \\

\begin{proposition}\label{grosse-proposition} In the setting of Theorem \ref{theo-ordre-1}, 
let ${\bf A}_n$ and ${\bf B}_n$ be uniformly bounded real diagonal
  matrices of size $N\times N$ and $n \times n$. Consider the following functions:
$$
\Phi({\bf Y})=\frac 1n
\tr \left( {\bf A} {\bf H} \frac{{\bf Y} {\bf B} {\bf Y}^*}{n} \right),\qquad 
\Psi({\bf Y})= \frac 1n \tr \left( {\bf A} {\bf H} {\bf D} {\bf H}
\frac{{\bf Y} {\bf B} {\bf Y}^*}{n} \right)\ .
$$
Then,
\begin{enumerate}
\item\label{variance-estimates} The following inequalities hold true:
$$
\mathrm{var}\left(\Phi({\bf Y})\right)= {\cal O}(n^{-2}),\qquad
\mathrm{var}\left(\Psi({\bf Y})\right)= {\cal O}(n^{-2})\ .
$$   
\item\label{approx-rules} The following approximations hold true:
\begin{eqnarray}
\EE\left[ {\Phi}({\bf Y}) \right] 
&=& 
\frac 1n \tr\left( \widetilde{\bf D} \widetilde{\bf T} {\bf B} \right) 
\frac 1n \tr\left( {\bf ADT} \right) 
+ {\cal O}\left( n^{-2} \right) \ , \label{approx-1}\\
\EE\left[ \Psi\left( {\bf Y} \right) \right]
&=& 
\frac{1}{1 - t^2 \gamma \tilde\gamma} 
\left(
\frac{1}{n^{2}} \tr\left( \widetilde{\bf D}\widetilde{\bf T} {\bf B} \right)
\tr\left( {\bf AD}^2{\bf T}^2 \right) 
-  
\frac{t \gamma}{n^{2}} \tr\left( \widetilde{\bf D}^2 \widetilde{\bf T}^2 {\bf B} \right)  \tr\left( {\bf ADT} \right) 
\right)
+ {\cal O}\left(\frac 1{n^2} \right) \ .\label{approx-2} 
\end{eqnarray}
\end{enumerate}
\end{proposition}

The variance inequalities are proved in Appendix \ref{appendix-variance}; 
the approximation rules, in Appendix \ref{appendix-approx}.

\section{First Order Moment Approximation: Proof of Theorem \ref{theo-ordre-1}} 
\label{sec-first-order} 
This section is devoted to the proof of the following approximation:
\begin{equation}
\label{eq-approx-ordre1}
\EE [ I_n(\rho) ] = V_n(\rho) + {\cal O}\left(n^{-1}\right) \ ,
\end{equation}
where 
\begin{equation}\label{valeur-V}
V_n(\rho) =  
\log \det \left( {\bf I} + \rho \delta_n(\rho) \widetilde{{\bf D}}_n \right) 
+ \log \det \left( {\bf I} + \rho \tilde{\delta}_n(\rho) {\bf D}_n \right) 
- n \rho \delta_n(\rho) \tilde{\delta}_n(\rho) \ . 
\end{equation}
This result already appears in \cite{mou-sim-sen-it03} and is proved
under greater generality in \cite{hac-lou-naj-(sub)aap05}. The proof presented here is new and relies on gaussian tools.

\subsection*{Outline of the proof}

The proof is divided into three steps. We first make some preliminary remarks.
Notice that the mutual information can be expressed as 
$I(\rho)= \int_0^{\rho} \tr\left( n^{-1} {\bf H}(t) {\bf YY}^*  \right) dt$. In particular,
\begin{equation}
\label{eq-forme-integrale}
\EE\left[ I(\rho) \right]
 = 
\int_{0}^{\rho} \tr \left( 
\EE\left[  {\bf H}(t) \frac{{\bf YY}^*}{n} \right]
\right) \, dt\ .
\end{equation}
In order to study the asymptotic behaviour of
$\EE\left[ I(\rho) \right]$, it is thus enough to study $ \tr\left(
  {\bf H}(t) \frac{{\bf YY}^*}{n} \right)$ for $n \rightarrow +\infty$
up to an integration. The Resolvent identity \eqref{eq-resolvent-id}
yields
$$
\tr\EE \left( {\bf H}(t) \frac{{\bf YY}^*}{n} \right) =\tr \EE \left(\frac{{\bf I}- {\bf H}(t)}t\right)\ .
$$
We are therefore led to the study of $\EE \left[ \tr ( {\bf H}(t) )
\right]$. We now describe the three steps of the proof.

\begin{enumerate}
\item[A.] In the first part of the proof, we expand $\EE {\bf H}(t)$ with
  the help of the Integration by parts formula
  \eqref{eq-integ-parts-our-model}. This derivations will bring to the
  fore the deterministic diagonal matrix ${\bf R}$, and Poincar\'e-Nash
  inequality will then allow us to obtain the following approximation:
$$
\EE \tr {\bf AH}  = \tr {\bf AR} + {\mathcal O}\left( n^{-1}\right) \ ,
$$
for every diagonal matrix ${\bf A}$ bounded in the spectral norm. Here
are the main steps, gathered in an informal way. Differentiating the
term $\EE \left( \left[ {\bf H} {\bf y}_j \right]_{p}
  \overline{Y_{pj}} \right)$, we obtain:
$$
\EE \left( \left[ {\bf H} {\bf y}_j \right]_{p} \overline{Y_{p,j}} \right) = 
d_p \tilde{d}_j \EE \left[ H_{pp} \right] -  t \tilde{d}_j
\EE \left( \frac{1}{n} \tr ({\bf D} {\bf H}) \, 
\left[ {\bf H} {\bf y}_j \right]_p 
\overline{Y_{pj}} \right) \ , 
$$
from which we will extract $\EE[ H_{pp}]$ later on. At this point,
Poincar\'e-Nash inequality yields some decorrelation up to ${\mathcal
  O}\left(n^{-1}\right)$ and we obtain:
$$
\EE \left[ \frac{1}{n} \tr ({\bf D} {\bf H}) \, ({\bf H} {\bf y}_j)_p 
\overline{Y_{pj}} \right]
\simeq
\EE  \left[ \frac{1}{n} \tr ({\bf D} {\bf H}) \right]  
\EE \left[ \left[ {\bf H} {\bf y}_j \right]_p \overline{Y_{pj}} \right]
= \alpha \EE \left[ \left[ {\bf H} {\bf y}_j \right]_p \overline{Y_{pj}} \right]
\ .
$$
This approximation allows us to isolate $\EE \left( \left[ {\bf H} {\bf y}_j
  \right]_{p} \overline{Y_{pj}} \right)$:
$$
( 1 + t \tilde{d}_j \alpha) \EE \left( \left[ {\bf H} {\bf y}_j \right]_{p} \overline{Y_{p,j}} \right)
\simeq d_p \tilde{d}_j \EE \left[ H_{pp} \right]\quad   \Leftrightarrow \quad 
\EE \left( \left[ {\bf H} {\bf y}_j \right]_{p} \overline{Y_{p,j}} \right) 
\simeq d_p \tilde{d}_j \tilde{r}_j \EE \left[ H_{pp} \right]\ .
$$
Now summing over $j$ and using the Resolvent identity $\EE H_{pp}  =  
1 - \frac{t}{n} \sum_{j=1}^{n} 
\EE \left[ {\bf H} {\bf y}_j \right]_{p} \overline{Y_{pj}}  $
in the previous equation yields:
$$
\frac{1-\EE H_{pp}}t \simeq \tilde \alpha d_p \EE H_{pp},\quad \textrm{that is}\quad  
\EE H_{pp} \simeq r_p\ .
$$
All the technical details are provided in Section \ref{first-step-outline-1}.\\

\item[B.] The second step follows from the approximation rule \eqref{approx-uno} stated 
in Section \ref{sec-cornerstone}, which immediatly yields
$$
\EE \tr {\bf AH}  = \tr {\bf AT} + {\mathcal O}\left( n^{-1} \right) \ .
$$ 
This in turn will imply that
$$
\EE \tr \left({\bf H}(t) \frac{{\bf Y}{\bf Y}^{*}}{n}
\right)  = \tr \left( \frac{{\bf I} -\EE{\bf H}}{t}\right) 
= \tr \left( \frac{{\bf I} -{\bf T}}t \right) +\varepsilon_n(t)
\stackrel{(a)}{=}
n \delta(t) \tilde{\delta}(t) + 
\varepsilon_n(t).
$$
where $(a)$ follows from the fact that ${\bf I-T}=t\tilde \delta {\bf D}({\bf I} + t\tilde \delta {\bf D})^{-1}$.\\
\item[C.] In the third step, we integrate the previous equality:
$$
\int_0^{\rho} \EE \tr \left({\bf H}(t) \frac{{\bf Y}{\bf Y}^{*}}{n}
\right) dt  = n \int_0^{\rho} \delta(t) \tilde{\delta}(t) dt + 
\int_0^{\rho} \varepsilon_n(t) dt.
$$
We identify $ n \int_0^{\rho} \delta(t) \tilde{\delta}(t) dt$ with $V_n(\rho)$ as given by \eqref{valeur-V},
and check that $\int_0^{\rho} \varepsilon_n(t) dt ={\mathcal O}(n^{-1})$.

\end{enumerate}

\subsection{Development of $\EE \left( \tr  {\bf A} {\bf H}(t)  \right)$
and Approximation by $\tr {\bf A} {\bf R}(t)$}\label{first-step-outline-1}

In order to study $\EE \left( \tr  {\bf A} {\bf H}(t) \right) $, we 
first consider the diagonal entries $H_{pp}(t)$ of 
${\bf H}(t)$. 
%The Resolvent identity (\ref{eq-resolvent-id}) immediately yields:
%\begin{equation}
%\label{eq-resolvente-id_pp}
%\EE\left[ H_{pp} \right] =  
%1 - \frac{1}{n} \sum_{j=1}^{n} 
%\EE \left( \left[ {\bf H} {\bf y}_j \right]_{p} \overline{Y_{pj}} \right) \ . 
%\end{equation}
For each index $j$, we have 
$$
\EE \left( \left[ {\bf H} {\bf y}_j \right]_{p} \overline{Y_{p,j}} \right) = 
\sum_{i=1}^{N} \EE \left( H_{pi} Y_{ij}  \overline{Y_{pj}} \right)  \ . 
$$
We now apply the Integration by parts formula
\eqref{eq-integ-parts-our-model} to the summand of the right hand side
for function $\Gamma$ defined as $\Gamma({\bf Y}) = H_{pi}
\overline{Y_{pj}}$. This yields:
\begin{equation}
\label{eq-utilisation-integ-parts-1}
\EE \left( H_{pi} Y_{ij}  \overline{Y_{pj}} \right)
 = 
d_i \tilde{d}_j \EE \left[ H_{ii} \right] \delta(i-p) 
- d_i \tilde{d}_j \frac{t}{n} 
\EE \left( \left[ {\bf H} {\bf y}_j \right]_p H_{ii} 
\overline{Y_{pj}} \right)  \ . 
\end{equation} 
Therefore, 
\begin{equation}
\label{eq-utilisation-integ-parts-2}
\EE \left( \left[ {\bf H} {\bf y}_j \right]_{p} \overline{Y_{p,j}} \right) = 
d_p \tilde{d}_j \EE \left[ H_{pp} \right] -  t \tilde{d}_j
\EE \left( \frac{1}{n} \tr ({\bf D} {\bf H}) \, 
\left[ {\bf H} {\bf y}_j \right]_p 
\overline{Y_{pj}} \right) \ , 
\end{equation}
from which we sahh extract $\EE[ H_{pp}]$ later on. 
Recall at this point that $\mathrm{var}\left(n^{-1} \tr {\bf D} {\bf
    H}(t) \right)={\mathcal O}\left(n^{-2}\right)$ by
Proposition \ref{PN-controle-1}. Recall also the following notations:
$
\beta = n^{-1} \tr ({\bf D} {\bf H}),$
$\alpha = \EE \left[ \beta \right],$  and $\overcirc{\beta}=\beta-\alpha.$
Plugging the relation $\beta = \alpha + \overcirc{\beta}$ into
(\ref{eq-utilisation-integ-parts-2}), 
we get   
\begin{equation}
\label{eq-utilisation-integ-parts-3}
\EE \left[ \left[ {\bf H} {\bf y}_j \right]_{p} \overline{Y_{p,j}} \right] 
= d_p \tilde{d}_j \EE [H_{pp}] 
-  t \tilde{d}_j \alpha 
\EE \left[ \left[ {\bf H} {\bf y}_j \right]_{p} \overline{Y_{p,j}} \right]
- t \tilde{d}_j 
\EE \left[ \overcirc{\beta} \, [{\bf H} {\bf y}_j]_p \overline{Y_{pj}} \right]\ .
\end{equation}
Solving this equation w.r.t. 
$\EE \left[ [{\bf H} {\bf y}_j]_{p} \overline{Y_{p,j}} \right]$ provides:
\begin{equation}
\label{eq-utilisation-integ-parts-4}
\EE \left[ [{\bf H} {\bf y}_j]_{p} \overline{Y_{p,j}} \right] 
= 
d_p \tilde{d}_j \tilde r_j  \EE [H_{pp}] 
- t \tilde{d}_j \tilde r_j  
\EE \left[ \overcirc{\beta} \, [{\bf H} {\bf y}_j]_p \overline{Y_{pj}} \right]
\quad\textrm{where}\quad 
\tilde{r}_{j}(t) = \frac{1}{1 + t \alpha(t) \tilde{d}_j} \quad 
\mathrm{for} \ 1\le j \le n  \ . 
\end{equation}
Summing (\ref{eq-utilisation-integ-parts-4}) over $j$ yields:
\begin{equation}
\label{eq-utilisation-integ-parts-5}
\EE  \left[{\bf H} \frac{{\bf Y} {\bf Y}^{*}}{n} \right]_{pp} 
= \tilde{\alpha} d_p \EE [ H_{pp} ]  
- t  \EE  \overcirc{\beta} \, 
\left[ {\bf H} \frac{{\bf Y} \tilde{{\bf D}} \tilde{{\bf R}} {\bf Y}^{*}}{n}
\right]_{pp} \ ,
\end{equation}
where $\widetilde{{\bf R}}$ is the diagonal matrix 
$\diag \left( \tilde{r}_j(t)\right) = 
\left({\bf I} + \alpha t \widetilde{\bf D} \right)^{-1}$ and 
$\tilde{\alpha} = \frac{1}{n} \tr \tilde{{\bf D}} \tilde{{\bf R}}$.
In order to obtain an expression for $\EE [H_{pp}]$, we plug the identity 
(\ref{eq-utilisation-integ-parts-5}) into the Resolvent identity:
\[
\EE [H_{pp}]  = 1 - 
t \EE \left[ \left[ {\bf H} \frac{{\bf Y} {\bf Y}^{*}}{n} \right]_{pp} 
\right] 
\]
and obtain:  
\begin{equation}
\label{eq-expre-Hpp}
\EE \left[ H_{pp} \right] = 
r_p + t^2 r_p 
\EE \left[ \overcirc{\beta} \, 
\left[ {\bf H} \frac{{\bf Y} \tilde{{\bf D}} \tilde{{\bf R}} {\bf Y}^{*}}{n}
\right]_{pp} 
\right]
\end{equation}
with
$
r_p(t) = \left(1 + t \tilde\alpha d_p\right)^{-1}. 
$
Let ${\bf A}$ be a $N\times N$ diagonal matrix with bounded spectral norm.
Multiplying \eqref{eq-expre-Hpp} by ${\bf A}$'s components and summing 
over $p$ yields:
$$
\EE \tr ({\bf A} {\bf H}) = 
\tr ({\bf A} {\bf R}) 
+ n t^{2} \EE \left[ \overcirc{\beta} \, \Phi({\bf Y}) \right] \ ,
$$
where 
$\Phi({\bf Y}) = \frac{1}{n} \tr  ({\bf A} {\bf R} {\bf H} \frac{{\bf Y} 
\widetilde{{\bf D}} \widetilde{{\bf R}} {\bf Y}^{*}}{n})$. 
As $\overcirc{\beta}$ is zero-mean, $ \EE[ \overcirc{\beta} \, \Phi ] = 
\EE [ \overcirc{\beta} \, \overcirc{\Phi} ]$. In particular, Cauchy-Schwarz inequality yields:
$$
|\EE  \overcirc{\beta} \, \overcirc{\Phi} |\le \sqrt{\mathrm{var}(\beta)}\sqrt{\mathrm{var}(\Phi)}. 
$$

Recall that $\mathrm{var}(\beta)={\mathcal O}\left(n^{-2}\right)$ by Prop.  \ref{PN-controle-1}. Since $\| {\bf R}_n
\|$ and $\| \widetilde{\bf D}_n \widetilde{\bf R}_n \|$ are both
bounded by Assumption {\bf (A1)} and by the definitions of ${\bf R}_n$
and $\widetilde{\bf R}_n$, one can directly apply the result of
Proposition \ref{grosse-proposition} to $\Phi$ in order to get
$\mathrm{var}(\Phi)={\mathcal O}\left(n^{-2}\right)$.

We have therefore proved the following: 
\vspace{0.03\columnwidth} 
\begin{proposition}
\label{prop-E(H)-R}
In the setting of Theorem \ref{theo-ordre-1}, let ${\bf A}$ be a
uniformly bounded diagonal $N \times N$ matrix.  Then for every $t\in \mathbb{R}^+$,
\begin{equation}
\label{eq-E(H)-R}
\EE (\tr {\bf A} {\bf H}(t)) = 
\tr {\bf A} {\bf R}(t) + {\cal O}\left(n^{-1}\right)\ .
\end{equation}
\end{proposition}
\vspace{0.03\columnwidth} 

\subsection{The Deterministic Approximation ${\bf T}(t)$.}  

Proposition \ref{prop-E(H)-R} provides a deterministic equivalent to
$\EE\, (\tr {\bf A} {\bf H})$ since matrix ${\bf R}$ is deterministic;
however its elements still depend on $\tilde{\alpha} = n^{-1} \tr
(\tilde{{\bf D}} \tilde{{\bf R}})$, which itself depends on $\alpha =
\EE \left( n^{-1} \tr {\bf D} {\bf H} \right)$, an unknown
parameter. The next step is therefore to apply Theorem \ref{theo-canonique}
to approximate matrix ${\bf R}$
by ${\bf T}$, which only depends on ${\bf D}$ and $\widetilde{\bf D}$ and 
and on $\delta$ and $\tilde\delta$, the solutions of \eqref{eq-equations-canoniques}.
Theorem \ref{theo-canonique} together with Equation (\ref{eq-E(H)-R}) imply that:
\begin{equation}
\label{eq-proximite-2}
\EE ( \tr {\bf A} {\bf H} ) = \tr ( {\bf A} {\bf T}) + 
{\cal O}\left(n^{-1}\right)\ .
\end{equation}
Since ${\bf T}$ only depends on $\delta$ and $\tilde\delta$, \eqref{eq-proximite-2} provides 
a deterministic equivalent of $\EE ( \tr {\bf A} {\bf H} )$ in terms of $\delta$ and $\tilde\delta$.
Note that taking ${\bf A}={\bf D}$ yields in particular
$\alpha = \delta + {\cal O}(n^{-2})$ 
while a direct application of Theorem \ref{theo-canonique} for 
$\widetilde{{\bf A}} = \widetilde{{\bf D}}$ yields
$ \tilde{\alpha} = \tilde{\delta} + {\cal O}(n^{-2})$. 

We are now in a position to describe the behaviour of 
$\EE\,\tr \left({\bf H}(t) \frac{{\bf Y}{\bf Y}^{*}}{n}\right)$
by using the Resolvent identity. From (\ref{eq-resolvent-id}) and 
(\ref{eq-proximite-2}), taking ${\bf A} = {\bf I}$, we immediately obtain:
\[
\EE \,  \tr 
\left({\bf H}(t) \frac{{\bf Y}{\bf Y}^{*}}{n} \right) 
=  
\frac{1}{t} \tr \left( {\bf I} - {\bf T}(t) \right) + 
{\cal O}\left(n^{-1}\right) \ . 
\]
As $ {\bf I} - {\bf T}(t) = ({\bf T}(t)^{-1} - {\bf I}) {\bf T}(t) = t\tilde{\delta}(t) {\bf D} {\bf T}(t)$,
we eventually get that 
\begin{equation}
\label{eq-exprederiveeI}
\EE \left[ \tr \left({\bf H}(t) \frac{{\bf Y}{\bf Y}^{*}}{n}
\right) \right] = n \delta(t) \tilde{\delta}(t) + 
\varepsilon_n(t),
\end{equation}
where the error $\varepsilon_n(t)$ is a ${\mathcal O}(n^{-1})$ term.

\subsection{Recovering the Deterministic Approximation $V(\rho)$ of 
$\EE[I(\rho)]$.}\label{section-first-order-epsilon} 
As mentionned previously, $\varepsilon_n(t)$ is a ${\mathcal
  O}(n^{-1})$ term, i.e. $|\varepsilon_n(t)|\le K_t\, n^{-1}$.  One can
easily keep track of $K_t$ in the derivations that lead to \eqref{eq-exprederiveeI} and prove that $K_t$
is bounded on the compact interval $[0,\rho]$.  In particular, $|
\varepsilon_n(t) | < K n^{-1} $ on the compact interval $[0, \rho]$ for some $K>0$.
The proof of this fact is omitted.

 As
$\varepsilon_n(t)$ is uniformly bounded on $[0,\rho]$, we have
$\left| \int_{0}^{\rho} \varepsilon_n(t) dt \right| = {\cal
  O}(n^{-1})$.  Therefore,
\[
\EE [ I(\rho) ] = 
\int_{0}^{\rho} n \delta(t) \tilde{\delta}(t) 
+ {\cal O}\left(n^{-1}\right) \ . 
\]
Consider now 
\[
V(\rho) = W\left(\rho, \delta(\rho), \tilde{\delta}(\rho) \right),
\]
where function $W(\rho, \delta, \tilde{\delta})$ is defined by 
\[
W\left(\rho, \delta, \tilde{\delta}\right) = 
\log \det \left( {\bf I} + \rho \delta \tilde{{\bf D}} \right) 
+ \log \det \left( {\bf I} + \rho \tilde{\delta} 
{\bf D} \right) - 
n \rho \delta \tilde{\delta} \ .
\]
One can easily check that:
$$
\frac{\partial W}{\partial \delta} =   \rho \left( 
\tr\left(\widetilde{{\bf D}}(I + \rho \delta \widetilde{{\bf D}})^{-1}
\right) - n \tilde{\delta} \right)  \qquad \textrm{and}\qquad 
\frac{\partial W}{\partial \tilde{\delta}} =   
\rho \left( \tr\left( {\bf D}(I + \rho \tilde{\delta} {\bf D})^{-1} \right)
- n \delta \right) \ .
$$
As the pair $(\delta(\rho), \tilde{\delta}(\rho))$ satisfies 
(\ref{eq-equations-canoniques}), 
the above partial derivatives evaluated at point 
$(\rho, \delta(\rho), \tilde{\delta}(\rho))$ are zero. 
Therefore, 
\begin{equation}
\label{eq-derivative-V} 
\frac{d V}{ d \rho} = \left( \frac{\partial W}{\partial \rho} 
\right)_{(\rho, \delta(\rho), \tilde{\delta}(\rho))} = 
n \delta(\rho) \tilde{\delta}(\rho) \  
\end{equation} 
which in turn implies (\ref{eq-expreI}). Theorem \ref{theo-ordre-1} is proved. \\ 

%In order to recover the approximate expression of $\EE[I(\rho)]$ 
%presented in \cite{mou-sim-sen-it03},

\begin{remark}[On the deterministic approximation ${\bf T}$]
  The deterministic approximation ${\bf T}$ can be used to approximate
  functionals of the eigenvalues of ${\bf Y Y^*}$ other that the
  mutual information $\log\det( \rho n^{-1} {\bf Y Y^*} +I)$ (see for instance 
  \cite{hac-lou-naj-(sub)aap05}). This relies on a specific representation of ${\bf T}$:
  The spectral theorem for Hermitian matrices yields the integral
  representation:
$$
\frac 1n \tr{\bf H}_n(z)= 
\int_0^\infty \frac{N_n(d\lambda)}{1+\lambda z }, \quad  z\in \CC \setminus \RR_- \ ,
$$
where $N_n$ represents the empirical distribution of the eigenvalues
of ${\bf Y Y^*}$. It can be shown that $n^{-1} \tr {\bf T}$ admits a
similar representation:
$$
\frac 1n \tr{\bf T}_n(z) = 
\int_0^\infty \frac{\pi(d\lambda)}{1+\lambda z}, \quad  z\in \CC \setminus \RR_-\ ,
$$
where $\pi$ is a probability measure. Finally, one can prove that $\int_0^\infty f(\lambda) N_n(d\lambda) - \int_0^\infty
f(\lambda) \pi_n(d\lambda)$ converges to zero almost
surely for every continuous bounded
function (see \cite{hac-lou-naj-(sub)aap05} for details). 
\end{remark}

\section{Second order Analysis: Proof of Theorem \ref{theo-clt}}
\label{sec-second-order-clt}

This section is devoted to the proof of the Central Limit Theorem:
$$
\sigma_n^{-1}(\rho) \left( I_n(\rho) - V_n(\rho)\right) \xrightarrow[n\rightarrow \infty]{\mathcal L} {\mathcal N}(0,1),
$$
where $\xrightarrow[]{\mathcal L}$ stands for the convergence in distribution.

\subsection*{Outline of the proof}

Denote by $\psi_n(u,\rho) = \EE\left[ e^{\mathbf{i} u \left( I_n(\rho)
      - V_n(\rho) \right)} \right]$ the characteristic function of
$I_n(\rho) - V_n(\rho)$.  The proof is based on the fact that in order
to establish the convergence (in distribution) of
$\sigma_n^{-1}(\rho) \left( I(\rho) - V(\rho) \right)$ towards
${\cal N}(0,1)$, it is sufficient to prove that:
$$
h_n(u) = \psi_n(u,\rho) - e^{-u^2 \sigma_n^2\left(\rho\right) / 2} 
\xrightarrow[n\to\infty]{} 0,\quad \forall u\in \RR\ . 
$$ 
In fact, recall by Proposition \ref{existence-variance} that the sequence $u/\sigma_n(\rho)$ belongs 
to a compact interval ${\cal K}_u$ since $\sigma_n(\rho)$ is bounded away from zero. 
If now $h_n(u) \to 0$ for every $u$, it converges uniformly to zero 
on the compact set ${\cal K}_u$ due to the continuity of $h_n$. 
Therefore, 
$$ 
h_n\left(\frac u{\sigma_n(\rho)}\right) = 
\EE \exp\left(\mathbf{i} u \frac{ I_n(\rho) - V_n (\rho)}{\sigma_n(\rho)} \right) - e^{-u^2/2}\xrightarrow[n\rightarrow\infty]{} 0,
$$ 
which proves the CLT.
% As $e^{-u^2/2}$ is the characteristic function of the Gaussian standard
% law, we have the desired result by the continuity theorem for characteristic
% functions. \\ 
The proof of the convergence of $h_n(u)$ towards zero is divided into two steps.
\begin{enumerate}
\item[A.] We first differentiate $\psi_n(u,t)$ with respect to $t$ in order to obtain a differential equation of the form:
\begin{equation}
\label{eq-dpsi/dt}
\frac{\partial \psi_n(u,t)}{\partial t} = - \frac{u^2}{2} 
\eta_n(t) \psi_n(u,t) + \varepsilon_n(u,t)\ .
\end{equation} 
In order to obtain the differential equation (\ref{eq-dpsi/dt}), we
first develop $\partial \psi / \partial t$ with the help of the
Integration by parts formula \eqref{eq-integ-parts-our-model}. We then
use Poincar\'e-Nash inequality to prove that relevant variances are of
order ${\mathcal O}(n^{-2})$.  This will enable us to decorrelate
various expectations, i.e. to express them as products of expectations
up to negligible terms. We shall then use the approximation rules stated
in Proposition \ref{grosse-proposition}
in Section \ref{more-approx} to deal with the obtained expectations.\\

\item[B.] 
The second step is devoted to identify the variance, that is to prove the identity
$$
\int_0^{\rho} \eta_n(t)\,dt = \sigma_n^2(\rho),
$$
where $\sigma_n^2$ is given by \eqref{definition-variance}, i.e. 
$\sigma^2(\rho)=-\log(1-\rho^2 \gamma(\rho)\tilde\gamma(\rho))$.  \\

\item[C.] The third step is devoted to the integration of
  \eqref{eq-dpsi/dt}.  Instead of directly integrating
  \eqref{eq-dpsi/dt}, we introduce
  $K_n(u,\rho)=\psi_n(u,\rho)e^{\frac{u^2}2 \sigma_n^2(\rho)}$ which
  satisfies the following differential equation:
\begin{equation}\label{edp-deux}
\frac{\partial K_n(u,t)}{\partial t} = \varepsilon_n(u,t) e^{\frac{u^2}2\sigma^2_n(t)}\ .
\end{equation}
Taking into account the obvious facts that $\psi_n(u,0) = 1$, $\sigma_n^2(0) = 0$ and therefore that 
$K_n(u,0)=1$, we shall obtain that 
$$
K_n(u,\rho)= 1 +\int_0^{\rho} \varepsilon_n(u,t) e^{\frac{u^2}2\sigma^2_n(t)}\, dt\ ,
$$
and prove that $\int_0^{\rho} \varepsilon_n(u,t) e^{\frac{u^2}2\sigma^2_n(t)}\, dt = {\mathcal O}(n^{-1})$.
This will yield in turn that:
$$
\psi_n(u,\rho)\quad =\quad  \left(1 +{\mathcal O}(n^{-1})\right) e^{-\frac{u^2}2 \sigma_n^2(\rho)}
\quad \stackrel{(a)}{=}\quad  e^{-\frac{u^2}2 \sigma_n^2(\rho)} + {\mathcal O}(n^{-1})\ .
$$ 
where $(a)$ follows from Proposition \ref{existence-variance}.
\end{enumerate} 
The theorem will then be proved.

\subsection{The differential equation  $\partial_t \psi_n = -\frac{u^2}2 \eta_n \psi_n +\varepsilon_n$ }

Recall that $ \psi_n(u,\rho)= \varphi_n(u,\rho)
e^{-\mathbf{i}uV_n(\rho)} $ where $\varphi(u, t) = \EE\left(
  e^{\mathbf{i} u I(t)} \right)$.  As $V_n'(t)= n
\delta(t) \tilde\delta(t)$ by \eqref{eq-derivative-V}, we obtain:
\begin{equation} 
\label{eq-dpsi/dt-fct(phi)}
\frac{\partial \psi(u,t)}{\partial t} = 
e^{-\mathbf{i} u V(t)} \frac{\partial \varphi(u,t)}{\partial t} 
 - \mathbf{i} u n \delta(t) \tilde\delta(t) \psi(u,t)  \ . 
\end{equation} 
Since $I'(t) = n^{-1} \tr {\bf H}(t) {\bf Y} {\bf Y}^*$ by
\eqref{eq-forme-integrale}, we have:
\begin{equation}
\label{eq-dphi/dt} 
\frac{\partial \varphi(u,t)}{\partial t} = 
\mathbf{i} u \ \EE \left[ \tr \left( {\bf H}(t) \frac{{\bf YY}^*}{n} \right)
e^{\mathbf{i} u I(t)} \right] = 
\frac{\mathbf{i} u}{n} \sum_{p,i=1}^N \sum_{j=1}^n 
\EE\left[
Y_{ij} H_{pi} \overline{Y_{pj}} e^{\mathbf{i} u I} \right] \ . 
\end{equation} 
Applying the Integration by parts formula
\eqref{eq-integ-parts-our-model} to $\EE\left[ Y_{ij} H_{pi}
  \overline{Y_{pj}} e^{\mathbf{i} u I} \right]$ (which can be written
$\EE\left( Y_{ij} \Gamma({\bf Y}) \right)$ for $\Gamma({\bf Y})=H_{pi}
  \overline{Y_{pj}} e^{\mathbf{i} u I}$) and using the differentiation formulas \eqref{eq-dH/dY^*} and
\eqref{eq-dI/dY^*} yields:
\begin{eqnarray} 
\EE\left[ Y_{ij} 
H_{pi} \overline{Y_{pj}} e^{\mathbf{i} u I} \right] 
&=&  
d_i \tilde d_j 
\EE\left[
\frac{\partial}{\partial \overline{Y_{ij}}} 
\left( H_{pi} \overline{Y_{pj}} e^{\mathbf{i} u I} \right) \right],  \nonumber \\ 
&=& - \ \frac tn d_i \tilde d_j  
\EE\left[ \left[ {\bf Hy}_j \right]_{p} H_{ii} \overline{Y_{pj}} 
e^{\mathbf{i} u I} \right]  
 + d_i \tilde d_j \delta(i-p) 
\EE\left[ H_{pi} e^{\mathbf{i} u I} \right]\nonumber  \\ 
& & + \ \frac{\mathbf{i} u t}{n} d_i \tilde d_j 
\EE\left[ H_{pi} \overline{Y_{pj}} \left[ {\bf Hy}_j \right]_{i}
e^{\mathbf{i} u I} \right]\ .
\label{eq-EYHYexp(uI)} 
\end{eqnarray} 
We now sum over index $i$ and obtain:
\begin{eqnarray*} 
\EE\left[ \left[ {\bf Hy}_j \right]_{p} 
\overline{Y_{pj}} e^{\mathbf{i} u I} \right] 
&=& 
- \ t \tilde d_j 
\EE\left[
\beta \,  
\left[ {\bf Hy}_j \right]_{p} \overline{Y_{pj}} e^{\mathbf{i} u I} \right] 
 + d_p \tilde d_j \EE\left[ H_{pp} e^{\mathbf{i} u I} \right]  \\
& & + \ \frac{\mathbf{i} u t}{n} \tilde d_j 
\EE\left[
\left[ {\bf HDHy}_j \right]_{p} \overline{Y_{pj}} e^{\mathbf{i} u I} \right]\ ,
\end{eqnarray*} 
where $\beta=n^{-1} \tr {\bf D} {\bf H}$. Writing $\beta= \overcirc\beta + \alpha$ 
yields:
\begin{eqnarray} 
(1+t\alpha \tilde d_j) \EE\left[ \left[ {\bf Hy}_j \right]_{p} 
\overline{Y_{pj}} e^{\mathbf{i} u I} \right] 
&=& 
- \ t \tilde d_j 
\EE\left[
% \frac 1n \overbrace{\tr \left( {\bf DH} \right)}^{\circ} \  
\overcirc{\beta} \, 
\left[ {\bf Hy}_j \right]_{p} \overline{Y_{pj}} e^{\mathbf{i} u I} \right] 
 + d_p \tilde d_j  \EE\left[ H_{pp} e^{\mathbf{i} u I} \right] 
\nonumber \\
& & + \ \frac{\mathbf{i} u t}{n} \tilde d_j 
\EE\left[
\left[ {\bf HDHy}_j \right]_{p} \overline{Y_{pj}} e^{\mathbf{i} u I} \right] \ . 
\label{eq-EHYYexp(uI)-2} 
\end{eqnarray} 
We now take into account that $\tilde r_j(t) =(1+t\alpha \tilde d_j)^{-1}$ and sum over $j$:
\begin{eqnarray} 
\EE\left[ \left[ {\bf HYY}^* \right]_{pp} e^{\mathbf{i} u I} \right] 
&=& 
- \ t 
\EE\left[
% \frac 1n \overbrace{\tr \left( {\bf DH} \right)}^{\circ} \  
\overcirc{\beta} \, 
\left[ {\bf HY} \widetilde{\bf D} \widetilde{\bf R} {\bf Y}^* 
\right]_{pp} e^{\mathbf{i} u I} \right] 
 + n \tilde\alpha d_p \EE\left[ H_{pp} e^{\mathbf{i} u I} \right] \nonumber \\
& & + \ \frac{\mathbf{i} u t}{n} 
\EE\left[
\left[ {\bf HDHY} \widetilde{\bf D} \widetilde{\bf R} {\bf Y}^* 
\right]_{pp} e^{\mathbf{i} u I} \right] \ . 
\label{eq-E[HYY]_pp} 
\end{eqnarray} 

By the Resolvent identity \eqref{eq-resolvent-id}, 
$\EE\left[ H_{pp} e^{\mathbf{i} u I} \right] = 
\EE\left[ e^{\mathbf{i} u I} \right] - \frac{t}{n} 
\EE\left[ \left[ {\bf HYY}^* \right]_{pp} e^{\mathbf{i} u I} \right]$. 
Replace now in (\ref{eq-E[HYY]_pp}), recall that 
$r_p(t) = \left( 1 + t \tilde\alpha(t) d_p \right)^{-1}$ 
and sum over $p$ to obtain:
\begin{eqnarray}
\EE \left[ \tr \left( {\bf H} \frac{{\bf YY}^*}{n} \right)
e^{\mathbf{i} u I} \right] &=&  
\tr\left( {\bf DR} \right) \tilde\alpha \EE\left[ e^{\mathbf{i} u I} \right] \nonumber \\
& & + \ \mathbf{i} u t 
\EE\left[ \frac 1n 
\tr\left(
{\bf RHDH} \frac{ {\bf Y} \widetilde{\bf D} \widetilde{\bf R} {\bf Y}^*}
{n} \right) 
e^{\mathbf{i} u I} \right] \nonumber \\
&\phantom{=}& - \ t \ 
\EE\left[
\overcirc\beta \, 
\tr\left( 
{\bf RH} \frac{{\bf Y} \widetilde{\bf D} \widetilde{\bf R} {\bf Y}^*}
{n} 
\right)  
e^{\mathbf{i} u I} \right] \nonumber \\  
& \stackrel{\triangle}{=}& \chi_1 +\chi_2 +\chi_3\  . 
\label{eq-trHYYexp(iuI)} 
\end{eqnarray} 
Thanks to Theorem \ref{theo-canonique}, 
\begin{equation} 
\label{eq-terme1} 
\chi_1\quad =\ \tr\left({\bf DR} \right) \tilde\alpha \EE\left[ e^{\mathbf{i} u I} \right] 
\quad =\ \tr\left({\bf DT} \right) \tilde\alpha \EE\left[ e^{\mathbf{i} u I} \right] + {\cal O}(n^{-1})  \quad = \
n \delta \tilde\delta \EE\left[ e^{\mathbf{i} u I} \right] 
+ {\cal O}(n^{-1}). 
\end{equation}

In order to deal with $\chi_2$, we apply the results of Proposition
\ref{grosse-proposition} related to $\Psi({\bf Y})$ in the particular
case where ${\bf A}= {\bf R}$ and ${\bf B}=\widetilde{\bf
  D}\widetilde{\bf R}$. In this case, $\chi_2$ writes 
$
\chi_2= \mathbf{i} u t \EE \left( \Psi({\bf Y}) e^{\mathbf{i} u I}\right),
$
and  Cauchy-Schwarz inequality yields:
$$
\left| \EE\left( \Psi e^{\mathbf{i} u I} \right) - 
\EE\left( e^{\mathbf{i} u I} \right) 
\EE\left( \Psi \right) \right| 
= 
\left| \EE[ e^{\mathbf{i} u I} \  \overcirc{\Psi} ] \right| 
\leq 
\sqrt{ \EE\left[ \left|\overcirc{\Psi}({\bf Y})\right|^2 \right] } 
={\cal O}(n^{-1})\ .
$$  
Therefore,
$$
\EE\left( \Psi e^{\mathbf{i} u I}\right) = \EE\left( e^{\mathbf{i} u I} \right) 
\EE\left( \Psi \right) +{\mathcal O}(n^{-1})\ .
$$
We now use the approximation for $\EE \Psi({\bf Y})$ 
given in Proposition \ref{grosse-proposition}. By Theorem \ref{theo-canonique}, we can replace
$\tilde{\bf R}$ (resp. $\widetilde {\bf R}$ by $\tilde{\bf T}$ (resp. $\widetilde{\bf T}$) 
in the obtained expression. We therefore obtain: 
\begin{eqnarray}
\EE\left( \Psi({\bf Y})e^{\mathbf{i} u I} \right)
&=&  
\EE\Psi({\bf Y})
\EE\left[ e^{\mathbf{i} u I} \right] + {\cal O}\left( n^{-1} \right)
\nonumber \\
&=& 
\frac{1}{1 - t^2 \gamma \tilde\gamma} 
\left(
\tilde\gamma 
\frac{1}{n} \tr\left( {\bf D}^2{\bf T}^3 \right) 
- t \gamma 
\frac{1}{n} \tr\left( \widetilde{\bf D}^3 \widetilde{\bf T}^3  \right) 
\frac{1}{n} \tr\left( {\bf DT}^2 \right) 
\right)
\EE\left[ e^{\mathbf{i} u I} \right] + {\cal O}\left(n^{-1} \right) \ . 
\label{eq-terme3} 
\end{eqnarray}

The term $\chi_3$ can be handled similarly: We apply the results of Proposition
\ref{grosse-proposition} related to $\Phi({\bf Y})$ in the particular
case where ${\bf A}= {\bf R}$ and ${\bf B}=\widetilde{\bf
  D}\widetilde{\bf R}$. In this case, $\chi_3$ writes 
$
\chi_2= -tn \EE \left(\overcirc{\beta} \Phi({\bf Y}) e^{\mathbf{i} u I}\right),
$
and  Cauchy-Schwarz inequality yields:
$$
\left| \EE\left( \overcirc{\beta} \Phi e^{\mathbf{i} u I} \right) - 
\EE\left( \overcirc\beta e^{\mathbf{i} u I} \right) 
\EE\left( \Phi \right) \right| 
= 
\left| \EE[ \overcirc\beta\, e^{\mathbf{i} u I} \  \overcirc{\Phi} ] \right| 
\leq 
\sqrt{ \EE\left[ {\overcirc\beta}^2 \right] } 
\sqrt{ \EE\left[ \overcirc{\Phi}^2 \right] } 
={\cal O}(n^{-2})\ .
$$  
We therefore obtain
\begin{eqnarray} 
\EE\left[
\overcirc{\beta} \, 
\tr\left( 
{\bf RH} \frac{{\bf Y} \widetilde{\bf D} \widetilde{\bf R} {\bf Y}^*}
{n} 
\right)  
e^{\mathbf{i} u I} \right] &=&  
\EE\left[
% \frac 1n \overbrace{\tr \left( {\bf DH} \right)}^{\circ} \  
\overcirc{\beta} \, 
e^{\mathbf{i} u I} \right] 
\ \tr\left( \widetilde{\bf D}^2 \widetilde{\bf T} \widetilde{\bf R} \right)
\frac{1}{n} \tr\left( {\bf DTR} \right) 
+ {\cal O}\left(n^{-1} \right)  \nonumber \\
&\stackrel{(a)}{=}&
\EE\left[
% \frac 1n \overbrace{\tr \left( {\bf DH} \right)}^{\circ} \  
\overcirc{\beta} \, 
e^{\mathbf{i} u I} \right] 
\  \tilde\gamma \  
\tr\left( {\bf DT}^2 \right) 
+ 
{\cal O}\left( n^{-1} \right)  \ ,
\label{eq-terme2} 
\end{eqnarray} 
where $(a)$ follows from Theorem \ref{theo-canonique}. It remains to
deal with the term $\EE\left[ \overcirc\beta \, e^{\mathbf{i} u I}
\right]$. To this end, we shall rely on \eqref{eq-E[HYY]_pp} and
develop the term $\EE\left[ H_{pp} e^{\mathbf{i} u I} \right]$.
The Resolvent identity yields:
$$
\EE\left[ \left[ {\bf HYY}^* \right]_{pp} e^{\mathbf{i} u I} \right] = \frac nt \EE\left[ e^{\mathbf{i} u I} \right] - \frac nt 
\EE\left[ H_{pp} e^{\mathbf{i} u I} \right] \ .
$$
Plugging this equality into \eqref{eq-E[HYY]_pp} and using $r_p = \left( 1 + t \tilde \alpha d_p \right)^{-1}$, 
we obtain after some computations 
\begin{eqnarray}
\EE\left[ 
\overcirc\beta \, 
e^{\mathbf{i} u I} \right] &=& 
t^2 
\EE\left[
\overcirc\beta \, 
e^{\mathbf{i} u I} \,  
\frac 1n \tr \left(
{\bf RDH}
\frac{{\bf Y}\widetilde{\bf D} \widetilde{\bf R} {\bf Y}^*}{n} \right)
\right] \nonumber \\ 
& & - \ \frac{\mathbf{i} u t^2}{n} 
\EE\left[
\frac 1n \tr \left(
{\bf RDHDH} 
\frac{ {\bf Y}\widetilde{\bf D} \widetilde{\bf R} {\bf Y}^*}{n} \right)
e^{\mathbf{i} u I} \right]  
+  
\frac{1}{n} \tr\left( {\bf D}\left( {\bf R} - \EE\left[{\bf H}\right]
\right)\right)
\EE\left[ e^{\mathbf{i} u I} \right] 
\nonumber \\
&\stackrel{(a)}{=}& t^2 \gamma \tilde\gamma \EE\left[ 
% \frac 1n \overbrace{\tr \left( {\bf DH} \right)}^{\circ}  \
\overcirc\beta \, 
e^{\mathbf{i} u I} \right] 
- \frac{1}{n} 
\frac{\mathbf{i} u t^2}{\left( 1 - t^2 \gamma \tilde\gamma \right)} 
\left( 
\tilde\gamma \frac 1n \tr\left( {\bf D}^3 {\bf T}^3 \right) 
- t \gamma^2 \frac 1n \tr\left( \widetilde{\bf D}^3 \widetilde{\bf T}^3 
\right) \right) \varphi 
+{\cal O}(n^{-2}) \label{eq-tr(DH)circ}
\end{eqnarray}  
where $(a)$ follows from Theorem \ref{theo-canonique}, Proposition \ref{grosse-proposition} 
and Proposition \ref{prop-E(H)-R}. We therefore obtain:
$$
\EE\left[ 
% \frac 1n \overbrace{\tr \left( {\bf DH} \right)}^{\circ} \
\overcirc\beta \, 
e^{\mathbf{i} u I} \right] = 
- \frac{1}{n} 
\frac{\mathbf{i} u t^2}{\left( 1 - t^2 \gamma \tilde\gamma \right)^2} 
\left( 
\tilde\gamma \frac 1n \tr\left( {\bf D}^3 {\bf T}^3 \right) 
- t \gamma^2 \frac 1n \tr\left( \widetilde{\bf D}^3 \widetilde{\bf T}^3 
\right) \right) \varphi 
+ {\cal O}\left( \frac{1}{n^2} \right)\ . 
$$
Plugging \eqref{eq-tr(DH)circ} into (\ref{eq-terme2}), and the result together with 
(\ref{eq-terme1}) and (\ref{eq-terme3}) into
(\ref{eq-trHYYexp(iuI)}), and getting back to (\ref{eq-dphi/dt}) and (\ref{eq-dpsi/dt-fct(phi)}),
we obtain:
$$
\frac{\partial \psi_n(u,t)}{\partial t} = - u^2 \eta_n(t) \psi_n(u,t) 
+ {\cal O}(n^{-1})\ , 
$$
where
\begin{equation} 
\label{eq-eta-brut} 
\eta_n(t) = \frac{1}{1 - t^2 \gamma \tilde\gamma} 
\left(
- \frac{t^2 \gamma  
\frac{1}{n} \tr\left( \widetilde{\bf D}^3 \widetilde{\bf T}^3 \right)
\frac{1}{n} \tr\left( {\bf DT}^2 \right)}
{1 - t^2 \gamma \tilde\gamma} 
+ t \tilde\gamma  
\frac{1}{n} \tr\left( {\bf D}^2{\bf T}^3 \right)
+ 
\frac{t^3 \tilde\gamma^2  
\frac{1}{n} \tr\left( {\bf D}^3 {\bf T}^3 \right)
\frac{1}{n} \tr\left( {\bf DT}^2 \right)}
{1 - t^2 \gamma \tilde\gamma} 
\right) \ .  
\end{equation} 
Equation (\ref{eq-dpsi/dt}) is established, and the first step of the proof is completed.

\subsection{Identification of the variance}

In order to finish the proof, it remains to prove that:
\begin{equation}\label{identification-variance}
\eta_n(t) = \frac 12 \frac{d\sigma_n^2(t)}{dt} \qquad \textrm{where} \quad
\sigma_n^2(t)= -\log\left( 1-t\gamma_n(t) \tilde\gamma_n(t)\right)\ . 
\end{equation}
To this end, we first begin by computing the derivatives of $\gamma_n(t)$ and $\tilde\gamma_n(t)$.
We shall prove that 
\begin{equation}\label{derive-gamma}
\frac{d \tilde\gamma}{dt} = 
- 2 \frac{   
\frac{1}{n} \tr\left( \widetilde{\bf D}^3 \widetilde{\bf T}^3 \right)
\frac{1}{n} \tr\left( {\bf DT}^2 \right)}
{1 - t^2 \gamma \tilde\gamma} \qquad \textrm{and}
\qquad  
\frac{d \gamma}{dt} = 
- 2 \frac{   
\frac{1}{n} \tr\left( {\bf D}^3 {\bf T}^3 \right)
\frac{1}{n} \tr\left( \widetilde{\bf D}\widetilde{\bf T}^2 \right)}
{1 - t^2 \gamma \tilde\gamma} \ .
\end{equation}
We only derive $\frac{d\tilde\gamma}{dt}$, the computations being similar in the other case.
We first expand the expression of $\tilde\gamma$, and obtain:
\begin{equation}
\label{eq-dtildegamma/dt-brut} 
\frac{d\tilde\gamma}{dt} = 
\frac{1}{n} \sum_{j=1}^n \tilde d_j^2 
\frac{d}{dt} \widetilde{T}_{jj}^2 = 
\frac{1}{n} \sum_{j=1}^n \tilde d_j^2 
\frac{d}{dt} \left( \frac{1}{1+ t \delta(t) \tilde d_j} \right)^2 
= 
- 2 \frac{d}{dt}\left( t \delta(t) \right) 
\frac{1}{n} \tr\left( \widetilde{\bf D}^3 \widetilde{\bf T}^3 \right) \ .
\end{equation} 
Let us now compute $\delta'(t)$:
\begin{equation}\label{delta-derivative}
\delta'(t) = 
\frac{1}{n} \sum_{i=1}^N d_i 
\left( \frac{1}{1+ t \tilde\delta(t)  d_i} \right)'  
=
- \gamma \tilde\delta(t) - \gamma t \tilde\delta'(t) \ . 
\end{equation}
A similar computation yields $\tilde\delta'(t) = 
- \tilde\gamma \delta(t) - \tilde\gamma t \delta'(t)$. Combining both equations yields:
$$
\delta' = \frac{ t \gamma \tilde\gamma \delta - \gamma \tilde\delta}
{ 1 - t^2 \gamma \tilde\gamma}\ .
$$ 
We now plug this into (\ref{eq-dtildegamma/dt-brut}) and obtain:
\begin{equation}\label{dernier-intermediaire}
\frac{d \tilde\gamma}{dt} = 
- 2 \frac{   
\frac{1}{n} \tr\left( \widetilde{\bf D}^3 \widetilde{\bf T}^3 \right)
\left( \delta - t \gamma \tilde\delta  \right)}
{1 - t^2 \gamma \tilde\gamma} \ . 
\end{equation}
Recall now that the mere definition of ${\bf T}$, $\widetilde{\bf T}$, $\delta$
and $\tilde \delta$ yields
\begin{equation}\label{eq-identities} 
\left\{
\begin{array}{l} 
t \delta \widetilde{\bf D} \widetilde{\bf T} = {\bf I} - \widetilde{\bf T} \\
t \tilde\delta {\bf D T} = {\bf I} - {\bf T} 
\end{array}\right. \ .
\end{equation} 
Using (\ref{eq-identities}), we obtain:
\begin{eqnarray} 
n^{-1} \tr\left( {\bf DT}^2 \right) &=& 
n^{-1} \tr\left( {\bf DT}\left( {\bf I} - t \tilde\delta 
{\bf DT} \right) \right) = \delta - t \tilde\delta \gamma \ , \label{eq-trDT2}\\
n^{-1} \tr\left( \widetilde{\bf D}\widetilde{\bf T}^2 \right) &=& 
n^{-1} \tr\left( \widetilde{\bf D}\widetilde{\bf T}\left( {\bf I} - t \delta 
\widetilde{\bf D}\widetilde{\bf T} \right) \right) = \tilde\delta - t \delta \tilde\gamma \ . \label{eq-trDT2-nontilde}
\end{eqnarray}
It remains to plug \eqref{eq-trDT2} in \eqref{dernier-intermediaire} to conclude the
proof of \eqref{derive-gamma}.

We are now in position to prove \eqref{identification-variance}. The
main idea in the following computations is to express
(\ref{eq-eta-brut}) as a symmetric quantity with respect to $\delta$
and ${\bf T}$ on the one hand and $\tilde\delta$ and $\tilde{\bf T}$
on the other hand. To this end, we split $\eta_n(t)$ in \eqref{eq-eta-brut}
as
$
\eta_n(t) = \frac{1}{1 - t^2 \gamma \tilde\gamma} \left( \eta^{(1)} + 
\eta^{(2)} + \eta^{(3)} \right)$. We first work on $\eta^{(3)}$:
\begin{eqnarray*}
\eta^{(3)} &\stackrel{(a)}{=}& \frac{t^3 \delta \tilde\gamma^2 \frac{1}{n} 
\tr\left( {\bf D}^3 {\bf T}^3 \right)}
{1 - t^2 \gamma \tilde\gamma} 
-
\frac{t^4 \tilde\delta \tilde\gamma^2 \gamma
\frac 1n \tr\left( {\bf D}^3 {\bf T}^3 \right)}
{1 - t^2 \gamma \tilde\gamma} \ ,\\
&\stackrel{(b)}{=}& \frac{- t^2 \tilde\gamma 
\frac{1}{n} \tr\left( {\bf D}^3 {\bf T}^3 \right)
\frac{1}{n} \tr\left( \widetilde{\bf D} \widetilde{\bf T}^2 \right) }
{1 - t^2 \gamma \tilde\gamma} 
+ 
t^2 \tilde\gamma \tilde\delta 
\frac{1}{n} \tr\left( {\bf D}^3 {\bf T}^3 \right) \ . 
\end{eqnarray*}
where $(a)$ follows from \eqref{eq-trDT2}, and $(b)$ from \eqref{eq-trDT2-nontilde}.
We now look at $\eta^{(2)}$:
$$
\eta^{(2)} + t^2 \tilde\gamma \tilde\delta
\frac{1}{n} \tr\left( {\bf D}^3 {\bf T}^3 \right) = 
 t \tilde\gamma  \left( 
\frac 1n \tr\left( {\bf D}^2 {\bf T}^3 + 
\frac{1}{n} \tr\left( {\bf D}^2 {\bf T}^2 
\left( t \tilde\delta {\bf DT} \right) \right) \right) \right) = 
 t \gamma \tilde\gamma
$$
where the last equality follows  (\ref{eq-identities}) again. We therefore have
\begin{eqnarray*}
\eta_n(t) &=& \frac{1}{1 - t^2 \gamma \tilde\gamma} 
\left(
- \frac{t^2 \gamma  
\frac{1}{n} \tr\left( \widetilde{\bf D}^3 \widetilde{\bf T}^3 \right)
\frac{1}{n} \tr\left( {\bf DT}^2 \right)}
{1 - t^2 \gamma \tilde\gamma} 
-
\frac{t^2 \tilde\gamma  
\frac{1}{n} \tr\left( {\bf D}^3 {\bf T}^3 \right)
\frac{1}{n} \tr\left( \widetilde{\bf D} \widetilde{\bf T}^2 \right)}
{1 - t^2 \gamma \tilde\gamma} 
+ t \gamma \tilde\gamma
\right)\ ,\\
&\stackrel{(a)}{=}&  \frac{1}{2} 
\frac{t^2 \gamma \tilde\gamma' + t^2 \gamma' \tilde\gamma + 2 t 
\gamma\tilde\gamma}{1 - t^2 \gamma \tilde\gamma} \ ,\\
&=& 
- \frac{1}{2} 
\frac{d}{dt} \log \left( 1 - t^2 \gamma \tilde\gamma \right)  \ ,
\end{eqnarray*}
where $(a)$ follows from \eqref{derive-gamma}. This concludes the identification of the variance.

\subsection{Integration of the differential equation \eqref{eq-dpsi/dt}}
Let us introduce $K_n(u,\rho)=\psi_n(u,\rho)e^{\frac{u^2}2
  \sigma_n^2(\rho)}$. Due to \eqref{eq-dpsi/dt}, $K_n(u,\rho)$ readily
satisfies the following differential equation:
\begin{equation}\label{eq-diff-K}
\frac{\partial K_n(u,t)}{\partial t} = \varepsilon_n(u,t) e^{\frac{u^2}2\sigma^2_n(t)}\ .
\end{equation}
As in Section \ref{section-first-order-epsilon}, one can easily prove
that $|\varepsilon_n(t)|\le \frac{K}n$ for every $t\in [0,\rho]$. As $K_n(u,0)=1$, we get
$$
K_n(u,\rho)= 1 +\int_0^{\rho} \varepsilon_n(u,t) e^{\frac{u^2}2\sigma^2_n(t)}\, dt\ .
$$
Due to Proposition \ref{definition-variance}, $\sigma_n^2(t)$ is bounded from above uniformly 
in $n$ and $t\in [0,\rho]$. This fact, together with $|\varepsilon_n(t)|\le \frac{K}n$ implies that:
$$
K_n(u,\rho)= 1 +{\mathcal O}\left(\frac 1n\right)\ .
$$
This in turn yields 
\begin{eqnarray*}
\Psi_n(u,\rho)&=& \left( 1 +{\mathcal O}\left(n^{-1}\right)\right) e^{-\frac {u^2}2 \sigma_n^2(\rho)}\\
&=& e^{-\frac {u^2}2 \sigma_n^2(\rho)} +{\mathcal O}(n^{-1})\ ,
\end{eqnarray*}  
where the last equality follows from the fact that $\sigma_n^2(\rho)$ is uniformly bounded by $n$ 
by Proposition \ref{existence-variance}.

\appendix
\subsection{Proof of Proposition \ref{existence-unicite}}\label{proof-existence-unicite}
Let us first establish the existence and uniqueness of the solution of 
(\ref{eq-equations-canoniques}). To this end, 
we plug the expression of $\tilde{\delta}$ in (\ref{eq-equations-canoniques}).
The system of two equations reduces to the single equation 
$\delta = f(t, \delta)$ where $f(t, \delta)$ is defined by  
\begin{equation}
\label{eq-single-canonique}
f(t,\delta) = \frac{1}{n} \tr \left( 
{\bf D} \left( {\bf I} 
+ t  \frac{1}{n} 
\tr \left( \widetilde{{\bf D}} 
\left({\bf I} + t \delta \widetilde{\bf D} \right)^{-1} 
\right) {\bf D} \right)^{-1} \right)
\end{equation}
which is itself equivalent to $g(\delta,t) = 1$ where
\[
g(t,\delta)  =  \frac{f(t,\delta)}{\delta} 
= \frac{1}{n} \tr \left( 
{\bf D} \left( \delta {\bf I} 
+ t  \frac{1}{n} \tr \left( \delta \widetilde{\bf D} 
\left({\bf I} + t \delta \widetilde{\bf D} \right)^{-1} \right) 
{\bf D} \right)^{-1} \right) \ . 
\]
The function $\delta \mapsto g(t,\delta)$ is continuous, decreasing
and satisfies $g(t,0) = +\infty$ and $g(t,+\infty) = 0$. Therefore, the
equation $g(t,\delta) = 1$ has a unique solution $\delta(t) > 0$.\\

The integral representation \eqref{representation-stieltjes} of $\delta$
and $\tilde\delta$ is related to the Stieltjes representation of a
class of analytic functions. One can indeed prove that functions $t
\mapsto \delta(t)$ and $t \mapsto \tilde{\delta}(t)$ defined on
$\RR^{+*}$, extend to $\CC \setminus \RR_-$, are analytic over this set
and satisfy the system (\ref{eq-equations-canoniques}) for every $z
\in \CC \setminus \RR_-$. Relying on specific properties of $\delta(z)$ and
$\tilde{\delta}(z)$, we can prove that the following integral representation holds:
\begin{equation}
\label{eq-representation-delta}
\delta(z) =  \int_{0}^{+\infty} \frac{ \mu(d\lambda)}{1 + \lambda z}  \qquad \textrm{and}\qquad 
\tilde{\delta}(z) =   \int_{0}^{+\infty} \frac{\tilde{\mu}(d\lambda)}{1 + \lambda z}  \ ,
\end{equation}
where $\mu$ and $\tilde{\mu}$ are nonnegative measures uniquely
defined on $\RR^+$ satisfying $\mu(\RR^+) = \frac{1}{n} \tr({\bf D})$
and $\tilde{\mu}(\RR^+) = \frac{1}{n} \tr(\widetilde{{\bf D}})$.
We refer to \cite{hac-lou-naj-(sub)aap05} where a more general result
is proven and skip the details.

\subsection{Proof of Proposition \ref{existence-variance}}\label{preuve-existence-variance}
In order to prove Proposition \ref{existence-variance}, it is sufficient to first prove that 
$1-t^2 \gamma \tilde \gamma$ is bounded away from zero and then to prove that the same quantity 
is strictly lower than 1, uniformly in $n$. We shall proceed into four steps.

\subsubsection{A priori estimates for $\delta$, $\tilde \delta$,
  $\gamma$ and $\tilde \gamma$} The mere definition of $\delta$ and $\tilde \delta$ yields:
\begin{equation}\label{upper-bound-delta}
\delta =\frac 1n \sum_{i=1}^N \frac{d_i}{1+td_i\tilde \delta} \le
\frac {N d_{\max}}n\quad \textrm{and}\quad 
\tilde \delta =\frac 1n \sum_{j=1}^n \frac{\tilde d_j}{1+t\tilde d_j \delta} \le \tilde d_{\max}\ .
\end{equation}
Using these upper estimates, one gets the following lower estimates:
\begin{equation}\label{lower-bound-delta}
\delta \ge \frac {\frac 1n  \tr\, {\bf D}}
{1+t d_{\max} \tilde d_{\max}}\quad \textrm{and}\quad 
\tilde \delta  \ge \frac{ \frac 1n \tr\, \widetilde{\bf D}}
{1+t\frac Nn d_{\max} \tilde d_{\max}}\ .
\end{equation}
One can notice that due to Assumption {\bf (A1)}, these lower bound are eventually bounded away from zero.
Finally a straightforward application of Jensen's inequality yields:
\begin{equation}\label{estimate-gamma}
\delta^2 =\left( \frac 1n \sum_{i=1}^N d_i T_{ii} \right)^2 \le \frac {N\gamma}n \quad  \textrm{i.e.} \quad 
\frac nN \delta^2 \le \gamma
\quad  \textrm{and} \quad \tilde \delta^2 \le \tilde \gamma\ .
\end{equation}

\subsubsection{An estimate over $\frac{d\tilde \delta}{dt}$}
The following equalities
are straightforward (see for instance \eqref{delta-derivative}):
\begin{equation}\label{derivatives-delta}
\delta'(t) = 
- \gamma \tilde\delta(t) - \gamma t \tilde\delta'(t)\quad \textrm{and}
\quad \tilde\delta'(t) = 
- \tilde\gamma \delta(t) - \tilde\gamma t \delta'(t)\ .
\end{equation}
In particular, $|\tilde \delta'(0)|= \tilde \gamma(0) \delta(0) \le Nn^{-1} \tilde d_{\max}^2 d_{\max}$
which is eventually bounded. Recall that $\tilde \delta$ admits the following representation:
$$
\tilde\delta(t)=\int_0^{\infty} \frac{\tilde\mu(d\lambda)}{1+t\lambda},
$$ 
where $\tilde \mu$ is a nonnegative mesure satisfying
$\tilde\mu(\mathbb{R}^+) =\frac 1n \tr\,\widetilde{\bf D}$. In particular, one obtains:
\begin{equation}\label{estimate-derivative}
0< -\tilde\delta'(t)\ =\ \int_0^{\infty} \frac{\lambda \tilde\mu(d\lambda)}{(1+t\lambda)^2}\  \le\  -\tilde \delta'(0) 
\ \le\  Nn^{-1} \tilde d_{\max}^2 d_{\max}\ .
\end{equation}
\subsubsection{The quantity $1-t^2 \gamma \tilde \gamma$ is bounded away from zero, uniformly in $n$ and for $t\in [0,\rho]$}
Eliminating $\delta'$ between the two equations in \eqref{derivatives-delta} yields:
$$
\frac{d\tilde \delta}{dt}(1-t^2 \gamma \tilde\gamma)\ =\ \tilde\gamma (t\tilde \delta \gamma - \delta) 
\ =\ \frac{\tilde \gamma}{n} \tr 
{\bf D} {\bf T} 
\left(t \tilde \delta {\bf D} {\bf T} - {\bf I} \right)\ =\ -\frac{\tilde \gamma}n \tr {\bf D} {\bf T}^2\ , 
$$
where the last equality follows from the identity
${\bf T} = ({\bf I} + t \tilde \delta {\bf D})^{-1}$ which yields $(t\tilde \delta {\bf D} {\bf T} -{\bf I})= -{\bf T}$.
Otherwise stated:
$$
1-t^2 \gamma \tilde\gamma =\frac{\tilde \gamma \tr {\bf D} {\bf T}^2}{n(- \tilde \delta'(t))}.
$$
This immediatly implies that $1-t^2 \gamma \tilde\gamma$ is positive.
In order to check that it is bounded away from zero uniformly in $n$, notice
first that $n^{-1} \tr {\bf D} {\bf T}^2 \ge d_{\max}^{-1} \gamma$.
Collecting now the previous estimates \eqref{estimate-gamma} 
and \eqref{estimate-derivative}, we obtain:
$$
1-t^2 \gamma \tilde\gamma \ge \frac{n^2}{N^2}\frac{\delta^2 \tilde\delta^2}{d_{\max}^{2}\tilde d_{\max}^{2}}.
$$
Using \eqref{lower-bound-delta} and Assumption {\bf (A1)}, we obtain
that $1-t^2\gamma\tilde \gamma$ is bounded away from zero, uniformly
in $n$ and for $t\in [0,\rho]$.\\
 
\subsubsection{The quantity $1-t^2 \gamma \tilde \gamma$ is strictly bounded above from 1, uniformly in $n$}
The inequalities \eqref{estimate-gamma} together with \eqref{lower-bound-delta}
yield:
$$
\sup_n\left( 1-t^2 \gamma \tilde \gamma\right)\  \le\ \sup_n \left( 1-t^2 \frac nN \delta^2 \tilde \delta^2\right)\ <\ 1\ .
$$ 
This completes the proof of Proposition \ref{existence-variance}.

\subsection{Proof of Poincar\'e-Nash inequality} 
\label{appendix-poincare}
The proof is borrowed from \cite{pas-umj05}. 
Recall that ${\bs \xi} = [ \xi_1, \ldots, \xi_M ]^T$ is 
a complex Gaussian random vector which law is determined by 
$$
\EE [ {\bs \xi} ] = {\bf  0}\ ,\quad \EE [ {\bs \xi} {\bs \xi}^T ] = 
{\bf 0}\quad \textrm{and}\quad 
\EE [ {\bs \xi}{\bs \xi}^* ] = {\bs \Xi}.
$$ 
Let $\Gamma=\Gamma(\xi_1,\cdots,\xi_M,\bar{\xi}_1,\cdots, \bar{\xi}_M)$ be a 
${\mathcal  C}^1$ 
complex function polynomially bounded together with its derivatives. 
We shall prove here Poincar\'e-Nash inequality % (\ref{eq-np}). 
 $$
 \mathrm{var}\left( {\Gamma}({\bs \xi}) \right) \leq 
 \EE \left[ \nabla_z \Gamma({\bs \xi})^T \ {\bs \Xi} \ 
 \overline{\nabla_z \Gamma({\bs \xi})} 
 \right] 
 +
 \EE \left[ \left( \nabla_{\overline{z}}  \Gamma({\bs \xi}) \right)^* 
 \ {\bs \Xi} \ 
 \nabla_{\overline{z}} \Gamma({\bs \xi}) 
 \right] 
 \ ,
 $$
 where $\nabla_z\Gamma = [ \partial\Gamma / \partial z_1, \ldots, 
 \partial\Gamma / \partial z_M ]^T$ and 
 $\nabla_{\overline{z}}\Gamma = [ \partial\Gamma / 
 \partial \overline{z_1}, \ldots, 
 \partial\Gamma / \partial \overline{z_M} ]^T$. \\ 

Let ${\bf y}$ and ${\bf z}$ be two $\CC^{2M}$-valued 
jointly Gaussian vectors (which parameters will be specified below). 
% determined by
% $\EE {\bf x}^{(i)} = {\bf 0}$, 
% $\EE \left[ {\bf x}^{(i)} {{\bf x}^{(i)}}^T \right]  = {\bf 0}$ and  
% $\EE \left[ {\bf x}^{(i)} {{\bf x}^{(i)}}^* \right]  = {\bf C}^{(i)}$ 
% for $i=1, 2$. \\
Consider the Gaussian vector ${\bf x}(t)=\sqrt{t} {\bf y} +\sqrt{1-t}
{\bf z}$ and let $\Upsilon : \CC^{2M} \rightarrow \CC$ be a given smooth 
function $\Upsilon = \Upsilon( z_1, \ldots, z_{2M}, \overline{z_1}, 
\ldots, \overline{z_{2M}} )$. Then 
$$
% \begin{equation}
% \label{eq-EUpsilon}
\EE \Upsilon\left({\bf y}\right) - 
\EE \Upsilon\left( {\bf z}\right) =
\int_0^1 \frac{d}{dt}\EE \Upsilon ({\bf x}(t))\, dt \ . 
% \end{equation} 
$$
Let 
$\nabla_z\Upsilon = [ \partial \Upsilon / \partial z_1, \ldots, 
\partial \Upsilon / \partial z_{2M} ]^T$ 
and
$\nabla_{\overline z}\Upsilon = [ 
\partial \Upsilon / \partial {\overline {z_1}}, \ldots, 
\partial \Upsilon / \partial {\overline {z_{2M}}} ]^T$. Then 
\begin{equation} 
\label{eq-integrand-np}
\frac{d}{dt}\EE \Upsilon ({\bf x}(t)) 
= 
\EE \left[ 
\left( \frac{{\bf y}}{2\sqrt{t}} - 
\frac{{\bf z}}{2\sqrt{1-t}}\right)^T 
\cdot 
\nabla_z \Upsilon ({\bf x}(t))
+
\left( \frac{{\bf y}}{2\sqrt{t}} - 
\frac{{\bf z}}{2\sqrt{1-t}}\right)^* 
\cdot
\nabla_{\overline{z}} \Upsilon({\bf x}(t))
\right]  \ . 
\end{equation}
At this point, assume that ${\bf y} = [ {\bf u}^T,{\bf u}^T ]^T$ 
and ${\bf z} = [ {\bf v}^T,{\bf w}^T ]^T$ where ${\bf u}$, ${\bf v}$ and 
${\bf w}$ are independent $\CC^M$-valued Gaussian vectors having the same 
law as ${\bs \xi}$. 
% Therefore we have 
% $$
% C^{(1)}=\left(
% \begin{array}{cc}
% C & C\\
% C & C\\
% \end{array}\right) \quad \textrm{and}\quad 
% C^{(2)}=\left(
% \begin{array}{cc}
% C & 0\\
% 0 & C\\
% \end{array}\right)\ .
% $$
Moreover, put  
$
\Upsilon({\bf x}(t)) = 
\Gamma({\bf x}_1(t))
\overline{\Gamma({\bf x}_2(t))} 
$
where ${\bf x}(t)$ is partitioned as ${\bf x}(t) = [ {\bf x}_1^T(t), 
{\bf x}_2^T(t) ]^T$. Then 
$$
% \begin{equation}
% \label{eq-var-gamma-upsilon} 
\textrm{var}(\Gamma({\bf u}))= 
\EE \Upsilon\left({\bf y}\right) - 
\EE \Upsilon\left( {\bf z}\right)
% \end{equation}
$$ 
which
leads us to consider the right hand side of Equation (\ref{eq-integrand-np}). 
The first term there (call it $\chi_1$) writes 
% \begin{eqnarray}
\begin{eqnarray} 
\chi_1 &=& \EE \left[ 
\left( \frac{{\bf y}}{2\sqrt{t}} - 
\frac{{\bf z}}{2\sqrt{1-t}}\right)^T 
\nabla_z \Upsilon ({\bf x}(t)) \right] \nonumber \\ 
&=&   
\frac{1}{2 \sqrt{t}} 
\EE\left[
\overline{\Gamma({\bf x}_2(t))} \ 
{\bf u}^T \nabla_z \Gamma({\bf x}_1(t)) 
+ 
\Gamma({\bf x}_1(t)) \ 
{\bf u}^T \nabla_z \overline{\Gamma({\bf x}_2(t))}  
\right]  \nonumber \\
& &   - 
\frac{1}{2 \sqrt{1-t}} 
\EE\left[
\overline {\Gamma({\bf x}_2(t))} \ 
{\bf v}^T \nabla_z \Gamma({\bf x}_1(t)) 
+ 
\Gamma({\bf x}_1(t)) \ 
{\bf w}^T \nabla_z 
\overline{\Gamma({\bf x}_2(t))}  
\right] \ .
\label{eq-derivee-upsilon} 
\end{eqnarray}
Let us process the term 
$\EE\left[
\overline{\Gamma({\bf x}_2(t))} \ 
{\bf u}^T \nabla_z  \Gamma({\bf x}_1(t)) \right]$. Writing 
${\bf u} = [ U_1, \ldots, U_M]^T$ and 
${\bf x}_i(t) = [ X_{i,1}, \ldots, X_{i,M} ]^T$ for $i=1,2$, we have by
the Integration by Parts Formula \eqref{eq-integ-parts} 
\begin{eqnarray*} 
\EE\left[
\overline{\Gamma({\bf x}_2(t))}  \ 
U_p 
\frac{\partial \Gamma({\bf x}_1(t))}{\partial X_{1,p}} 
\right]
&=& 
\sum_{m=1}^M [ {\bs \Xi} ]_{pm} 
\EE \left[
\frac{\partial}{\partial \overline{U_m} } 
\left(
\overline{\Gamma({\bf x}_2(t))}  
\frac{\partial \Gamma({\bf x}_1(t))}{\partial X_{1,p}} 
\right) \right] \\
&=&
\sqrt{t} \sum_{m=1}^M [ {\bs \Xi} ]_{pm} 
\EE\left[ 
\frac{\partial \Gamma({\bf x}_1(t))}{\partial X_{1,p}} 
\frac{\partial \overline{\Gamma({\bf x}_2(t))}}{\partial \overline{X_{2,m}}} 
+
\overline{\Gamma({\bf x}_2(t))}  
\frac{\partial^2 \Gamma({\bf x}_1(t))}
{\partial X_{1,p} \partial \overline{X_{1,m}}} 
\right]
\end{eqnarray*} 
where we used ${\bf x}_1(t) = \sqrt{t} {\bf u} + \sqrt{1-t} {\bf v}$ and
${\bf x}_2(t) = \sqrt{t} {\bf u} + \sqrt{1-t} {\bf w}$ in the second 
equality. 
By treating similarly the other terms of the right hand side of
(\ref{eq-derivee-upsilon}) and taking the sum, the terms with the second 
order derivatives 
$\partial^2 / \partial X_{i,p} \partial \overline{X_{i,m}}$ disappear and we 
end up with
\begin{equation}
\label{eq-chi_1-np} 
% \EE \left[ 
% \left( \frac{{\bf y}}{2\sqrt{t}} - 
% \frac{{\bf z}}{2\sqrt{1-t}}\right)^T 
% \nabla_z \Upsilon ({\bf x}(t)) \right] 
\chi_1 
= \frac 12 
\EE \left[ 
\left( \nabla_z \Gamma({\bf x}_1(t)) \right)^T \ {\bs \Xi} \ 
\overline{\nabla_{z} \Gamma({\bf x}_2(t))}  
+ 
\left( \nabla_{\overline{z}} {\Gamma({\bf x}_2(t))} \right)^* 
\ {\bs \Xi} \ 
\nabla_{\overline{z}} {\Gamma({\bf x}_1(t))}  
\right] 
\end{equation} 
where we used the identity $\overline{\partial f / \partial z} = 
\partial \overline{f} / \partial \overline{z}$ which proof is straightforward. 
\\ 
By using twice the Cauchy-Schwarz inequality we obtain:
 \begin{eqnarray*} 
 \EE \left| 
\nabla_z \Gamma({\bf x}_1(t))^T \ {\bs \Xi} \ 
\overline{\nabla_z \Gamma({\bf x}_2(t))}  
\right|   
&\le&  
\EE \left[ \left( 
\nabla_z \Gamma({\bf x}_1(t))^T \ {\bs \Xi} \ 
\overline{\nabla_z \Gamma({\bf x}_1(t))} \right)^{\frac 12} \right. \\
& & 
\ \ \ \ \ \ \ \ \ \ \ \ 
\ \ \ \ \ \ \ \ \ \ \ \ 
\left. \left( 
\nabla_z \Gamma({\bf x}_2(t))^T \ {\bs \Xi} \ 
\overline{\nabla_z \Gamma({\bf x}_2(t))} \right)^{\frac 12}  
\right] \\
&\le& 
\left\{ \EE 
\left[
\nabla_z \Gamma({\bf x}_1(t))^T \ {\bs \Xi} \ 
\overline{\nabla_z \Gamma({\bf x}_1(t))} 
\right] \right\}^{\frac 12} \\
& & 
\ \ \ \ \ \ \ \ \ \ \ \ 
\ \ \ \ \ \ \ \ \ \ \ \ 
\left\{ \EE 
\left[ 
\nabla_z \Gamma({\bf x}_2(t))^T \ {\bs \Xi} \ 
\overline{\nabla_z \Gamma({\bf x}_2(t))} 
\right] \right\}^{\frac 12} \ . 
\end{eqnarray*} 
The second term of the right hand side of (\ref{eq-chi_1-np}) can be bounded
in a similar manner. 
Noticing that ${\bf x}_1(t)$ and ${\bf x}_2(t)$ have the same 
law as ${\bf u}$, which does not depend on $t$, it results that 
$$
| \chi_1 | \leq 
\frac{1}{2}
 \EE \left[ \nabla_z \Gamma({\bf u})^T \ {\bs \Xi} \ 
 \overline{\nabla_z \Gamma({\bf u})} 
 \right] 
 +
\frac 12  \EE \left[ \left( \nabla_{\overline{z}}  \Gamma({\bf u}) \right)^* 
 \ {\bs \Xi} \ 
 \nabla_{\overline{z}} \Gamma({\bf u}) 
 \right] \ . 
$$
The second term of the right hand side of Equation (\ref{eq-integrand-np})
is treated similarly, which leads to the desired result. 
% Recalling Equations (\ref{eq-EUpsilon}--\ref{eq-var-gamma-upsilon}) we obtain
% the desired result. 

\subsection{Proof of Theorem \ref{theo-canonique}}
\label{anx-proof-theo-canonique} 
We first give a sketch of the proof to emphasize the main ideas over the technical aspects of the proof.
\begin{enumerate}
\item We first prove that the asymptotic behaviour of $n^{-1}
  \tr\left( {\bf A}\left( {\bf R} - {\bf T} \right)\right)$ is
  directly related to the behaviour of $\alpha(t) - \delta(t)$.
  Similarly, $n^{-1} \tr \widetilde{\bf A}\left( \widetilde{\bf
        R} - \widetilde{\bf T} \right)$ is related to
  $\tilde\alpha(t) - \tilde\delta(t)$.
\item We extend the definition of $\alpha$ from $t\in \mathbb{R}^+$ to $z\in \CC \setminus \RR_-$ and
establish an integral representation:
$$
\alpha(t)=\int_{\mathbb{R}^+} \frac{\nu(d\lambda)}{1+\lambda t}\ .
$$
As a consequence of the integral representations for $\delta$, $\tilde
\delta$ and $\alpha$, we prove that $\delta$, $\tilde \delta$ and
$\alpha$ are bounded analytic functions on every compact subset of
$\CC \setminus \RR_-$.
\item As a consequence of this detour in the complex plane, we prove
  the following weaker result. For every uniformly bounded diagonal matrix 
  ${\bf A}$, the following holds true: 
$$
\left\{
\begin{array}{ll}
n^{-1} \tr ({\bf A} {\bf R}) & =  n^{-1} \tr ({\bf A} {\bf T}) + o(1)  \\
n^{-1} \tr (\tilde{{\bf A}} \tilde{{\bf R}}) & =  n^{-1} \tr (\tilde{{\bf A}} \tilde{{\bf T}}) + 
o(1) 
\end{array}\right. \ .
$$

\item We then refine the previous result in order the get the
  sharper rate of convergence ${\mathcal O}(n^{-2})$ instead of $o(1)$.
\end{enumerate}

The theorem will then be proved.

\subsubsection{The asymptotic behaviour of $n^{-1}
  \tr\left( {\bf A}\left( {\bf R} - {\bf T} \right)\right)$ and its relation with $\alpha(t) - \delta(t)$}
The standard matrix identity 
$$
{\bf R} - {\bf T} = {\bf R} ({\bf T}^{-1} - {\bf R}^{-1}) {\bf T}
$$
immediatly yields
\begin{eqnarray*}
n^{-1} \tr \left( {\bf A}({\bf R} - {\bf T}) \right) &=& 
t (\tilde{\delta}(t) -\tilde{\alpha}(t)) 
\frac{1}{n} \tr \left( {\bf A} {\bf R} {\bf D} {\bf T} \right) \quad  \textrm{and}\\
n^{-1} \tr \left( 
\widetilde{{\bf D}} (\widetilde{{\bf T}} - \widetilde{{\bf R}}) \right) &=& 
\tilde{\delta}(t) -\tilde{\alpha}(t) = 
t (\alpha(t) - \delta(t))  
\frac{1}{n} \tr \left( \widetilde{{\bf D}} \widetilde{{\bf R}} 
\widetilde{{\bf D}} \widetilde{{\bf T}} \right) \ .
\end{eqnarray*}
Therefore, 
\begin{equation}
\label{eq-calcul-intermediaire-1}
n^{-1} \tr \left( {\bf A}({\bf R} - {\bf T}) \right) = 
t^{2}  (\alpha(t) - \delta(t)) 
\frac{1}{n} 
\tr \left( \widetilde{{\bf D}} \widetilde{{\bf R}} \widetilde{{\bf D}} 
\widetilde{{\bf T}} \right)
\frac{1}{n} \tr \left( {\bf A} {\bf R} {\bf D} {\bf T} \right) \ .
\end{equation}

\subsubsection{An integral representation for $\alpha$, and bounds over $\alpha$, $\delta$ and $\tilde \delta$} 
Recall that $\alpha(t) = \EE[ n^{-1} \tr( {\bf D} ({\bf I} + t
n^{-1}{\bf Y} {\bf Y}^{*})^{-1})]$. This function readily extends from $t\in \RR^+$ to $z\in \CC\setminus \RR^-$.
Moreover, the following representation holds true:
\begin{equation}
\label{eq-representation-alpha}
\alpha(z) =  \int_{0}^{+\infty} \frac{\nu(d\lambda)}{1 + \lambda z} \ ,
\end{equation}
where  $\nu$ is a uniquely defined positive measure on $\RR^+$ such that 
$\nu(\RR^+) = \frac{1}{n} \tr {\bf D}$. To prove this, we introduce the 
eigenvalue/eigenvector decomposition of 
matrix $n^{-1}{\bf Y}{\bf Y}^{*} = \sum_{i=1}^{N} \lambda_i {\bf u}_i {\bf u}_i^{*}$
where $(\lambda_i,\ 1\le i\le N)$ and $({\bf u}_i,\ 1\le i\le N)$ represent its eigenvalues 
and eigenvectors respectively. The random variable 
$\beta(z) =  \frac{1}{n} \tr {\bf D} ({\bf I} + z \frac{{\bf Y}{\bf Y}^{*}}{n})^{-1}$  
can be written as
$$
\beta(z) = \frac{1}{n} \sum_{i=1}^{N} \frac{{\bf u}_i^{*} {\bf D} {\bf u}_i}{\lambda_i - z}
=\int_{0}^{+\infty} \frac{\omega(d\lambda)}{1 + \lambda z}  \ ,
$$
where $\omega$ is the nonnegative random measure defined by 
$$
\omega = \frac{1}{n} \sum_{i=1}^{N} {\bf u}_i^{*} {\bf D} {\bf u}_i \delta(\lambda-\lambda_i)\ .
$$ 
Consider now the measure $\nu$ defined by $\nu = \EE [\omega]$, that is $\nu(B) = \EE [\omega(B)]$
for every Borel set $B \subset \RR^+$. 
It is clear that $\alpha(z) = \EE[\beta(z)]$ is given by (\ref{eq-representation-alpha}), 
and that $\nu(\RR^+) = \EE[ \omega(\RR^+)]$ is given by 
\[
\nu(\RR^+) = \EE\left[
 \frac{1}{n} \sum_{i=1}^{N} {\bf u}_i^{*} {\bf D} {\bf u}_i \right] = 
\EE\left[ \frac{1}{n} \tr {\bf D} (\sum_i {\bf u}_i {\bf u}_i^{*}) \right]\ .
\]
As $\sum_i {\bf u}_i {\bf u}_i^{*} = {\bf I}$, 
$\nu(\RR^+) = \frac{1}{n} \tr {\bf D}$ as expected and representation 
(\ref{eq-representation-alpha}) implies that $\alpha(z)$ is analytic over $\CC\setminus \RR^-$.

Let $\mbox{dist}(w,\RR^+)$ stand for the distance from 
element $w \in \CC$ to $\RR^+$. Then the following holds true for every $z\in \CC\setminus \RR^-$:
\begin{equation}
\label{eq-bornitude-alpha}
|\alpha(z)| \leq \frac 1n \tr ({\bf D}) \frac 1{|z|} \frac{1}{\mbox{dist}(-\frac{1}{z}, 
\RR^+ )}\le \frac Nn
d_{\mathrm{max}} \frac{1}{|z|} \frac{1}{\mbox{dist}(-\frac{1}{z}, 
\RR^+ )} \ . 
\end{equation}    
Similarly, (\ref{eq-representation-delta}) yields that 
\begin{equation}
\label{eq-bornitude-delta}
|\delta(z)|
\leq \frac{Nd_{\mathrm{max}}}{n|z|} \frac{1}{\mbox{dist}(-\frac{1}{z}, \RR^+ )}\ .
\end{equation}
A similar result holds for $\tilde{\delta}_n(z)$. These upper bounds imply in particular that
$\alpha(z)$, $\delta(z)$ and $\tilde\delta(z)$ are uniformly bounded on each compact subset of 
$\CC \setminus \RR_-$.\\
%\emph{i.e.}, that for each compact subset ${\cal K} \subset \CC -
%\RR_-$, it exists a constant $K$, independent on $n$ (but depending on
%${\cal K}$) such that $\sup_{n} \sup_{z \in {\cal K}} |\delta_n(z)|
%\leq K$. A similar result
%holds for $\tilde{\delta}_n(z)$. \\

\subsubsection{A weaker result as a consequence of Montel's theorem}
We first establish that for every diagonal matrix ${\bf A}$ uniformly bounded,  
\begin{equation}
\label{eq-proximite-faible}
\left\{
\begin{array}{ll}
n^{-1} \tr ({\bf A} {\bf R}) & =  n^{-1} \tr ({\bf A} {\bf T}) + o(1)  \\
n^{-1} \tr (\tilde{{\bf A}} \tilde{{\bf R}}) & =  n^{-1} \tr (\tilde{{\bf A}} \tilde{{\bf T}}) + 
o(1) 
\end{array}\right. \ .
\end{equation}
We take \eqref{eq-calcul-intermediaire-1} as a starting point.
Matrices ${\bf R}, \widetilde{{\bf R}}, {\bf T}$, and $\widetilde{{\bf T}}$ 
have their spectral norms bounded by one for $t \in \RR^+$ and matrices 
${\bf A}, {\bf D}$, and $\widetilde{{\bf D}}$ are also uniformly bounded by
assumption. 
Therefore, the terms 
$n^{-1} \tr \left( \widetilde{{\bf A}} \widetilde{{\bf R}} 
\widetilde{{\bf D}} \widetilde{{\bf T}} \right)$ and 
$n^{-1} \tr \left( {\bf A} {\bf R} {\bf D} {\bf T} \right)$ are 
also bounded. In order to prove (\ref{eq-proximite-faible}), 
it is sufficient to prove that $\alpha(t) - \delta(t) = o(1)$. To this
end, we make use of Proposition \ref{prop-E(H)-R}
and write $\alpha(t) - \delta(t)$ as
\[
\alpha(t) - \delta(t)= \frac{1}{n} \tr \left( {\bf D}({\bf R} - {\bf T}) 
\right) + \varepsilon_n(t)  \ , 
\]
where $\varepsilon_n(t)={\mathcal O}(n^{-2})\ .$ Using relation (\ref{eq-calcul-intermediaire-1}) for ${\bf A} = {\bf D}$, 
we immediately get that: 
\begin{equation}
\label{eq-expre-alpha-delta}
\alpha(t) - \delta(t) = (\alpha(t) - \delta(t))  
t^{2} \frac{1}{n} 
\tr \left( \widetilde{{\bf D}} \widetilde{{\bf R}} \widetilde{{\bf D}} 
\widetilde{{\bf T}} \right) 
\frac{1}{n} \tr \left( {\bf D} {\bf R} {\bf D} {\bf T} \right) 
+ \varepsilon_n(t)\ .
\end{equation}

As $\sup_{n} \left( \|{\bf R}_n\|, \|\widetilde{{\bf R}}_n\|, \|{\bf T}_n\|, 
\|\widetilde{{\bf T}}_n\| \right) \leq 1$, we have:
\[
\frac{1}{n} 
\tr \left( \widetilde{{\bf D}} \widetilde{{\bf R}} \widetilde{{\bf D}} 
 \widetilde{{\bf T}} \right) 
\frac{1}{n} \tr \left( {\bf D} {\bf R} {\bf D} {\bf T} \right) 
\leq \frac{N}{n} d_{\mathrm{max}}^{2} \tilde{d}^{2}_{\mathrm{max}} 
\leq 2 c  d_{\mathrm{max}}^{2} \tilde{d}^{2}_{\mathrm{max}}
\]
as soon as $\frac{N}{n} \leq 2c$. Therefore, if 
$t < t_0 := (2 d_{\mathrm{max}} \tilde{d}_{\mathrm{max}} \sqrt{c})^{-1}$, 
then 
\[
t^{2} \frac{1}{n} 
\tr \left( \widetilde{{\bf D}} \widetilde{{\bf R}} \widetilde{{\bf D}}
\widetilde{\bf T} \right)
\frac{1}{n} \tr \left( {\bf D} {\bf R} {\bf D} {\bf T} \right) 
< \frac{1}{2}
\]
for $n$ large enough. Eq. (\ref{eq-expre-alpha-delta}) thus implies 
that  
\begin{equation}\label{convergence-petit-t}
|\alpha_n(t) - \delta_n(t)| < 2 |\varepsilon_n(t)|,\quad i.e. \quad \alpha(t) - \delta(t)={\mathcal O}(n^{-2}) \quad 
\textrm{for} \quad
t < t_0\ .
\end{equation}
This in particular implies that $\alpha_n(t) - \delta_n(t) = o(1)$ for $t
< t_0$; however, it remains to establish this convergence for $t \geq
t_0$. To this end, observe that $\alpha_n(z) -\delta_n(z)$ is analytic
in $\CC \setminus \RR_-$ and bounded on each compact subset of $\CC
\setminus \RR_-$. Montel's theorem asserts that the sequence of
functions $\alpha_n(z) -\delta_n(z)$ is compact and therefore that
there exists a converging subsequence which converges towards an
analytic function. Since this limiting function is zero on $[0,t_0[$
by \eqref{convergence-petit-t}, it must be zero everywhere due to the
analycity. Therefore from every subsequence, one can extract a
subsequence that converges toward zero. Necessarily, $\alpha_n(z)
-\delta_n(z)$ converges to zero for every $z\in \CC \setminus \RR^-$
and in particular for
$t \geq 0$. This establishes (\ref{eq-proximite-faible}). \\
Even if the convergence rate of $\alpha_n(t)-\delta_n(t)$ is ${\mathcal O}(n^{-2})$ for 
$t<t_0$, Montel's theorem does not imply that the convergence
rate of $\alpha_n(z) - \delta_n(z)$ remains ${\cal O}(n^{-2})$
elsewhere. Therefore, there remains some work to be done
in order to prove that
$\alpha_n(t) - \delta_n(t) = {\cal O}(n^{-2})$ for each $t > 0$. \\

\subsubsection{End of the proof}
We take (\ref{eq-expre-alpha-delta}) as a starting point.
Equations (\ref{eq-proximite-faible}) imply that for each $t\ge0$,   
\begin{equation}
\label{eq-proximite-faible-2}
\left\{
\begin{array}{ll}
n^{-1}  \tr \left( {\bf D} {\bf R}(t) {\bf D} {\bf T}(t) \right)
 - \gamma(t) & =  o(1) \\
n^{-1} \tr\left( \widetilde{{\bf D}} \widetilde{{\bf R}}(t) \widetilde{{\bf D}} 
\widetilde{{\bf T}}(t) \right) - \tilde{\gamma}(t) & =  o(1) 
\end{array}\right. \ .
\end{equation}
where $\gamma_n=n^{-1}\tr {\bf D}^2 {\bf T}^2$ and $\tilde\gamma_n=n^{-1}\tr \widetilde{\bf D}^2 \widetilde{\bf T}^2$.
Thanks to Proposition \ref{controle-variance}, (\ref{eq-proximite-faible-2}) implies that 
\[
\inf_n  \left(
1 - t^{2} \frac{1}{n} \tr \left( {\bf D}_n {\bf R}_n(t) {\bf D}_n {\bf T}_n(t) \right)
 \frac{1}{n} 
\tr\left( \widetilde{{\bf D}}_n \widetilde{{\bf R}}_n(t) \widetilde{{\bf D}}_n 
\widetilde{{\bf T}}_n(t) \right) \right)  > 0  \ . 
\]
Equation (\ref{eq-expre-alpha-delta}) thus clearly implies that
$\alpha(t) - \delta(t)$ is of the same order of magnitude as
$\varepsilon_n(t)$, i.e. that $\alpha(t) - \delta(t) = {\cal
  O}(n^{-2})$. Theorem \ref{theo-canonique} is proved.

\subsection{Proof of Proposition \ref{grosse-proposition}-\eqref{variance-estimates} - Variance controls}
\label{appendix-variance}
Consider first $\Phi({\bf Y})=\frac 1n
\tr \left( {\bf A} {\bf H} \frac{{\bf Y} {\bf B} {\bf Y}^*}{n} \right)$.
We use Poincar\'e-Nash inequality (\ref{eq-np-our-model}) to control the 
variance of $\Phi$. It writes 
\begin{equation}
\label{eq-np-grosse-prop} 
\EE \left[ \overcirc{\Phi}({\bf Y})^2 \right] 
\leq  
\sum_{i=1}^N \sum_{j=1}^n d_i \tilde d_j 
\EE \left[ \left| \frac{\partial \Phi}{\partial Y_{i,j}} \right|^2 \right] 
+
\sum_{i=1}^N \sum_{j=1}^n d_i \tilde d_j 
\EE \left[ \left| \frac{\partial \Phi}{\partial \overline{Y_{i,j}}} \right|^2 
\right] \ . 
\end{equation} 
We have $\Phi({\bf Y}) = (1/n^2) \sum_{p,r=1}^N \sum_{q=1}^n  
a_p b_q H_{pr} Y_{rq} \overline{Y_{pq}}$. From the differentiation
formula (\ref{eq-dH/dY}) we have
$$
\frac{\partial}{\partial Y_{ij}} \left( H_{pr} Y_{rq} \overline{Y_{pq}} 
\right) 
= - \frac tn H_{pi} [ {\bf y}_j^* {\bf H} ]_{r} Y_{rq} \overline{Y_{pq}} + 
H_{pr} \overline{Y_{pq}} \delta(r-i) \delta(q-j) \ . 
$$
Therefore, after a straightforward computation we obtain 
$\partial \Phi / \partial Y_{ij} = \phi^{(1)}_{ij} + \phi^{(2)}_{ij}$ with
$$
\phi^{(1)}_{ij} = - \frac{t}{n^3} \left[ {\bf y}_j^* {\bf HYB} {\bf Y}^*
{\bf AH} \right]_{i} 
\quad \mathrm{and} \quad 
\phi^{(2)}_{ij} = \frac{1}{n^2} b_j \left[ {\bf y}_j^* 
{\bf AH} \right]_{i} \ . 
$$
The first term of the right hand side of inequality (\ref{eq-np-grosse-prop}) 
can be treated as follows: 
\begin{eqnarray}
% \EE \left[ \overcirc{\Phi}({\bf Y})^2 \right] 
\sum_{i=1}^N \sum_{j=1}^n d_i \tilde d_j 
\EE \left[ \left| \frac{\partial \Phi}{\partial Y_{i,j}} \right|^2 \right] 
&\leq& 
2 \sum_{i=1}^N \sum_{j=1}^n d_i \tilde d_j 
\left( \EE\left[ \left| \phi^{(1)}_{ij} \right|^2 \right] + 
\EE \left[ \left| \phi^{(2)}_{ij} \right|^2 \right] \right) \nonumber \\
&=&
\frac{2 t^2}{n^6} \EE \left[
\tr\left( {\bf H YBY}^* {\bf AHDHAYBY}^* {\bf HY} \widetilde{\bf D} {\bf Y}^*  
\right) \right] \nonumber \\
& &  
+ \frac{2}{n^4} \EE\left[\tr\left(
{\bf AHDHAY} {\bf B}^2 \widetilde{\bf D} {\bf Y}^* 
\right)\right]  \ . 
\label{eq-np-lemme-trAHYBY} 
\end{eqnarray} 
Let $A = \sup\| {\bf A}_n \|$. Using inequalities 
(\ref{eq-tr(ab)<tr(aa)tr(bb)}), (\ref{eq-tr(ab)<|b|tr(a)}), (\ref{eq-|H|<1})
and Cauchy-Schwarz inequality, we have
\begin{eqnarray} 
\frac{2 t^2}{n^6} \EE \left[
\tr\left( 
{\bf H \ YBY}^* \ {\bf AHDHA\ YBY}^* \ {\bf H \ Y} \widetilde{\bf D} {\bf Y}^*  
\right) \right]   
\! \! \!  \! \! \!  \! \! \!  \! \! \! 
\! \! \!  \! \! \!  \! \! \!  \! \! \! 
\! \! \!  \! \! \!  \! \! \!  \! \! \! 
\! \! \!  \! \! \!  \! \! \!  \! \! \! 
\! \! \!  \! \! \!  \! \! \!  \! \! \! 
\! \! \!  \! \! \!  \! \! \!  \! \! \!  & & \nonumber \\ 
&\leq& 
\frac{2 t^2}{n^6} \EE\left[
\sqrt{ \tr\left( 
\left( {\bf H YBY}^* {\bf AHDHAYBY}^* {\bf H} \right)^2 
\right) }
\sqrt{
\tr\left(
\left({\bf Y} \widetilde{\bf D} {\bf Y}^* \right)^2
\right) 
} \right]  \nonumber \\
&\leq& 
\frac{2 t^2}{n^6} \EE\left[
\| {\bf H} \|^4 \| {\bf A} \|^2 \| {\bf D} \| 
\sqrt{ \tr\left( \left( {\bf YBY}^* \right)^4 \right) } 
\sqrt{
\tr\left(
\left({\bf Y} \widetilde{\bf D} {\bf Y}^* \right)^2
\right) 
} \right]  \nonumber \\
&\leq& 
\frac{2 d_{\mathrm{max}} A^2 t^2}{n^2} 
\sqrt{
\frac 1n \EE \left[ \tr\left( 
\left( \frac{{\bf YBY}^*}{n} \right)^4 \right) \right] } 
\sqrt{
\frac 1n \EE\left[ \tr\left( 
\left( \frac{{\bf Y}\widetilde{\bf D} {\bf Y}^*}{n} \right)^2 \right) \right]} 
\nonumber \\ 
&<& \frac{K}{n^2} \ , 
\label{eq-ineq-lemme-trAHYBY} 
\end{eqnarray} 
where the last inequality is due to (\ref{eq-trace-bounded}). 
% and to the assumed bounded character of $\| {\bf B}_n \|$ and 
% $\| \widetilde{\bf D}_n \|$. \\
Turning to the second term of the right hand side of 
(\ref{eq-np-lemme-trAHYBY}), we have
\begin{equation} 
\label{eq-ineq-lemme-trAHYBY-2}
\frac{2}{n^4} \EE\left[\tr\left(
{\bf AHDHAY} {\bf B}^2 \widetilde{\bf D} {\bf Y}^* 
\right)\right] 
\leq 
\frac{2 A^2 d_{\mathrm{max}}}{n^2} 
\EE\left[ \frac 1n \tr\left( \frac 1n {\bf Y B}^2 \widetilde{\bf D} {\bf Y}^*
\right) \right] < \frac{K}{n^2}\ .   
\end{equation} 
The second term of the right hand side of Inequality (\ref{eq-np-grosse-prop})
is treated similarly. 
This proves that $\mathrm{var}(\Phi)={\mathcal O}(n^{-2})$. \\ 

Consider now 
$\Psi({\bf Y})= \frac 1n \tr \left( {\bf A} {\bf H} {\bf D} {\bf H}
\frac{{\bf Y} {\bf B} {\bf Y}^*}{n} \right)$. 
The proof being quite similar to the previous one, 
we just give its main steps. 
By (\ref{eq-np-our-model}) we have 
$\EE [ \overcirc{\Psi}({\bf Y})^2 ] 
\leq  
\sum_{i=1}^N \sum_{j=1}^n d_i \tilde d_j \left(
\EE [ | \partial \Psi / \partial Y_{i,j} |^2 ] +
\EE [ | \partial \Psi / \partial \overline{Y_{i,j}} |^2 ] \right)$. 
A computation similar to above yields 
$\partial \Psi / \partial Y_{ij} = \psi^{(1)}_{ij} + \psi^{(2)}_{ij} 
+ \psi^{(3)}_{ij}$ where
\begin{eqnarray*} 
\psi^{(1)}_{ij} &=&  
- \frac{t}{n^3} 
\left[ {\bf y}_j^* {\bf HDHYBY}^* {\bf AH} \right]_{i}, \\  
\psi^{(2)}_{ij} &=&  
- \frac{t}{n^3} 
\left[ {\bf y}_j^* {\bf HYBY}^* {\bf AHDH} \right]_{i}, \\ 
\psi^{(3)}_{ij} &=&  
\frac{1}{n^2} b_j \left[ {\bf y}_j^* {\bf AHDH} \right]_{i} \ . 
\end{eqnarray*} 
We have 
\begin{eqnarray*}
\sum_{i=1}^N \sum_{j=1}^n d_i \tilde d_j 
\EE \left[ \left| \frac{\partial \Psi}{\partial Y_{i,j}} \right|^2 \right] 
&\leq&  
3 \sum_{i=1}^N \sum_{j=1}^n d_i \tilde d_j 
\left( \EE \left[ \left| \psi^{(1)}_{ij} \right|^2 \right] 
+ \EE \left[ \left| \psi^{(2)}_{ij} \right|^2 \right] 
+ \EE \left[ \left| \psi^{(3)}_{ij} \right|^2 \right] \right) \\ 
&=& \phantom{+} \frac{3 t^2}{n^6} 
\EE\left[\tr\left(
{\bf HDH} \ {\bf YBY}^* \ {\bf AHDHA\ YBY}^* \ {\bf HDH\ Y}
\widetilde{\bf D} {\bf Y}^* 
\right)\right] \\ 
& & 
+ \frac{3 t^2}{n^6} 
\EE\left[\tr\left(
{\bf H \ YBY}^* \   
{\bf A} \left({\bf HD}\right)^3 {\bf HA \ YBY}^* \ {\bf H \ Y}
\widetilde{\bf D} {\bf Y}^* \right)\right] \\ 
& & + \frac{3}{n^4} 
\EE\left[\tr\left(
{\bf A} \left({\bf HD}\right)^3 {\bf HA \ YB}^2 \widetilde{\bf D} {\bf Y}^* 
\right)\right] \ . 
\end{eqnarray*} 
The first two terms of the right hand side can be bounded by a series
of inequalities similar to inequalities (\ref{eq-ineq-lemme-trAHYBY}). 
The third term can be bounded as in (\ref{eq-ineq-lemme-trAHYBY-2}). This
ends the proofs of the variance controls in Proposition \ref{grosse-proposition}.

\subsection{Proof of Proposition \ref{grosse-proposition}-\eqref{approx-rules} - Approximation rules}
\label{appendix-approx}
Consider first $\Phi(Y)=\frac 1n
\tr \left( {\bf A} {\bf H} \frac{{\bf Y} {\bf B} {\bf Y}^*}{n} \right)$.
we write $\Phi({\bf Y}) = (1 / n^2) \sum_{p,i=1}^N \sum_{j=1}^n 
a_p b_j \EE\left[ Y_{ij} {H}_{pi} \overline{Y_{pj}} \right]$ and 
apply the Integration by parts formula \eqref{eq-integ-parts-our-model} to the summand.
Using identity (\ref{eq-dH/dY^*}), we have 
$$
\EE\left[ Y_{ij} H_{pi} \overline{Y_{pj}} \right] = 
d_i \tilde d_j \EE\left[ \frac{\partial}{\partial \overline{Y_{ij}}}
\left( H_{pi} \overline{Y_{pj}} \right)\right] = 
- \frac tn d_i \tilde d_j 
\EE\left[ \left[ {\bf Hy}_j \right]_{p} H_{ii} \overline{Y_{pj}} \right] 
+ d_i \tilde d_j \delta(i-p) \EE\left[ H_{pi} \right]  \ . 
$$
By taking the sum over the index $i$, we obtain
$\EE\left[ \left[ {\bf Hy}_j \right]_{p} \overline{Y_{pj}} \right] = 
- t \tilde d_j 
\EE\left[ 
% \frac 1n \tr\left({\bf DH}\right)
\beta \, 
\left[ {\bf Hy}_j \right]_{p} \overline{Y_{pj}} \right] + 
d_p \tilde d_j \EE\left[ H_{pp} \right]$. 
Writing now 
$\beta = \overcirc\beta + \alpha$ 
and then grouping together the terms with 
$\EE\left[ \left[ {\bf Hy}_j \right]_{p} \overline{Y_{pj}} \right]$,  
% and by recalling that $\tilde r_j(t) = 
% \left( 1 + t \alpha(t) \tilde d_j \right)^{-1}$, 
we obtain: 
$$
\EE\left[ \left[ {\bf Hy}_j \right]_{p} \overline{Y_{pj}} \right] = 
- t \tilde d_j \tilde r_j 
\EE\left[ 
% \frac 1n \overbrace{\tr\left({\bf DH}\right)}^{\circ} 
\overcirc\beta \, 
\left[ {\bf Hy}_j \right]_{p} \overline{Y_{pj}} \right] + 
d_p \tilde d_j \tilde r_j \EE\left[ H_{pp} \right]\ .
$$
We now sum over $j$ and $p$, and obtain:
$$
\EE\left[ \frac 1n \tr\left( {\bf AH} \frac{{\bf YBY}^*}{n} \right) \right] = 
\frac 1n \tr\left( \widetilde{\bf D} \widetilde{\bf R} {\bf B} \right)
\frac 1n \tr\left( {\bf AD} \ \EE\left[{\bf H}\right] \right) 
+ \varepsilon \ ,
$$
with 
$$
\varepsilon = 
- t \ 
\EE\left[ 
% \frac 1n \overbrace{\tr\left({\bf DH}\right)}^{\circ} 
\overcirc\beta \, 
\frac{1}{n} \tr\left(
{\bf AH} \frac{{\bf Y}\widetilde{\bf D}\widetilde{\bf R} {\bf BY}^*}
{n} \right) \right]
= 
- t \ 
\EE\left[ 
% \frac 1n \overbrace{\tr\left({\bf DH}\right)}^{\circ} 
  \overcirc\beta \, \frac{1}{n} \overbrace{\tr\left( {\bf AH}
      \frac{{\bf Y}\widetilde{\bf D}\widetilde{\bf R} {\bf BY}^*} {n}
    \right)}^{\circ} \right]\ .
$$  
Applying Cauchy-Schwarz inequality, Proposition \ref{PN-controle-1} and the
variance controls in Proposition \ref{grosse-proposition}, we get $|
\varepsilon | ={\mathcal O}(n^{-2})$.

By Theorem \ref{theo-canonique}, $n^{-1}\tr\left( \widetilde{\bf D}
  \widetilde{\bf R} {\bf B} \right)  = n^{-1} \tr\left( \widetilde{\bf D}
  \widetilde{\bf T} {\bf B} \right) + {\cal O}(n^{-2})$. By Theorem \ref{theo-canonique} 
and Proposition \ref{prop-E(H)-R}, we obtain 
$n^{-1} \tr\left( {\bf AD} \ \EE\left[{\bf H}\right] \right) = n^{-1} \tr\left(
  {\bf ADT} \right)  + {\cal O}(n^{-2})$. This ends the proof of \eqref{approx-1}.

Consider now $\Psi({\bf Y})= \frac 1n \tr \left( {\bf A} {\bf H} {\bf
    D} {\bf H} \frac{{\bf Y} {\bf B} {\bf Y}^*}{n} \right).$ In order
to compute $\mathbb{E} \Psi({\bf Y})$, we shall need the following
intermediate result:

\begin{lemma}\label{lm-trDHDH} 
In the setting of Theorem \ref{theo-ordre-1}, let $\Upsilon({\bf Y}) = \frac{1}{n} \tr\left( {\bf DHDH}\right) $. 
Then 
\begin{enumerate}
\item \label{eq-var-trDHDH} 
The following estimate holds true:
\begin{equation*}
\mathrm{var}\left[ \Upsilon({\bf Y}) \right] ={\mathcal O}\left(\frac 1{n^2}\right)\ ,
\end{equation*}
\item \label{eq-mean-trDHDH} moreover,  
$$
\EE\left[ \Upsilon({\bf Y}) \right] = 
\frac{\gamma}{1 - t^2 \gamma \tilde\gamma} + 
{\cal O}\left( \frac{1}{n^2} \right) \ . 
$$
\end{enumerate}
\end{lemma}
\vspace{0.03\columnwidth} 

\begin{proof}
  In order to prove Lemma \ref{lm-trDHDH}-(\ref{eq-var-trDHDH}), we
  use the Resolvent identity (\ref{eq-resolvent-id}) and write:
$$
{\bf DHDH} = {\bf DHD} - tn^{-1} {\bf DHDHYY}^*\ .
$$
Since $\mathrm{var}(X+Y)\le 2 \mathrm{var}(X) + 2\mathrm{var}(Y)$, 
we only need to deal with
each term of the right handside. By Proposition \ref{PN-controle-1},
$\mathrm{var} (n^{-1} \tr\, {\bf DHD})={\mathcal O}(n^{-2})$ and by
Proposition \ref{grosse-proposition}-\eqref{variance-estimates},
$\mathrm{var} (tn^{-2} \tr\, {\bf DHDHYY}^*)={\mathcal O}(n^{-2})$ and
the proof of Lemma \ref{lm-trDHDH}-(\ref{eq-var-trDHDH}) is completed.
\\
Let us now prove Lemma \ref{lm-trDHDH}-(\ref{eq-mean-trDHDH}). The Resolvent identity
(\ref{eq-resolvent-id}) yields:
\begin{equation} 
\label{eq-HDH_pp}
\EE\left[ \left[ {\bf HDH} \right]_{pp} \right] = 
d_p \EE\left[ H_{pp} \right] - t 
\EE\left[ \left[ {\bf HDH} \frac{{\bf YY}^*}{n} \right]_{pp} \right] \ .
\end{equation} 
We then write 
$\EE\left[ \left[ {\bf HDH} \frac{{\bf YY}^*}{n} \right]_{pp} 
\right] = n^{-1}
\sum_{k,i=1}^N \sum_{j=1}^n d_k H_{pk} H_{ki} Y_{ij} \overline{Y_{pj}}$,  
and apply the differentiation formula (\ref{eq-dH/dY}) to the summand. 
After derivations similar to 
(\ref{eq-EYHYexp(uI)}--\ref{eq-EHYYexp(uI)-2}), we obtain:
\begin{eqnarray}
\frac{1}{n} \EE\left[ \left[ {\bf HDHy}_j \right]_{p} 
\overline{Y_{pj}} \right]
&=& 
- \frac{t}{n} \tilde d_j \tilde r_j 
\EE\left[ \left[ {\bf Hy}_j \right]_{p} \overline{Y_{pj}} 
\frac{1}{n} \tr\left( {\bf DHDH} \right) \right] \nonumber \\
& & 
- \frac{t}{n} \tilde d_j \tilde r_j 
\EE\left[
% \frac{1}{n} \overbrace{\tr\left( {\bf DH} \right)}^{\circ} 
\overcirc\beta \, 
\left[ {\bf HDHy}_j \right]_{p} \overline{Y_{pj}} \right] \nonumber \\
& &
+ \frac{1}{n} d_p \tilde d_j \tilde r_j 
\EE\left[ \left[ {\bf HDH} \right]_{pp} \right] \ . 
\label{eq-HDHYpj)} 
\end{eqnarray} 
% \begin{eqnarray} 
% \EE\left[ \left[ {\bf HDH} \frac{{\bf YY}^*}{n} \right]_{pp} \right] 
% &=& -t \ \EE\left[ 
% \left[ {\bf H} \frac{{\bf Y} \widetilde{\bf D} \widetilde{\bf R}
% {\bf Y}^*}{n} \right]_{pp} 
% \frac{1}{n} \tr\left( {\bf DHDH} \right) \right] \nonumber \\ 
% & & -t \  \ \EE\left[
% \frac{1}{n} \overbrace{\tr\left( {\bf DH} \right)}^{\circ} 
% \left[ {\bf HDH} \frac{{\bf Y} \widetilde{\bf D} \widetilde{\bf R}
% {\bf Y}^*}{n} \right]_{pp} \right] \nonumber  \\ 
% & & + \tilde\alpha d_p \EE\left[ \left[ {\bf HDH} \right]_{pp} \right] \ . 
% \label{eq-HDHYY*} 
% \end{eqnarray} 
Taking the sum over $j$ and combining with (\ref{eq-HDH_pp}) yields: 
\begin{eqnarray}
\EE\left[ \left[ {\bf HDH} \right]_{pp} \right] 
&=&
\phantom{+} t^2 r_p \EE\left[
\left[ {\bf H} \frac{{\bf Y} \widetilde{\bf D} \widetilde{\bf R}
{\bf Y}^*}{n} \right]_{pp} 
\frac{1}{n} \tr\left( {\bf DHDH} \right) \right] \nonumber \\
& & 
+ t^2 r_p \EE\left[
% \frac{1}{n} \overbrace{\tr\left( {\bf DH} \right)}^{\circ} 
\overcirc\beta \, 
\left[ {\bf HDH} \frac{{\bf Y} \widetilde{\bf D} \widetilde{\bf R}
{\bf Y}^*}{n} \right]_{pp} \right]  \nonumber  \\ 
& & 
+ r_p d_p \EE\left[ H_{pp} \right]  \ . 
\label{eq-E(HDH)_pp} 
\end{eqnarray} 
Taking now the sum over $p$, we obtain:
\begin{equation}
\label{eq-trDHDH-1} 
\EE\left[ \frac{1}{n} \tr\left( {\bf DHDH} \right) \right] 
= 
\frac{1}{n} \sum_{p=1}^N d_p 
\EE\left[ \left[ {\bf HDH} \right]_{pp} \right]  = 
\chi_1 + \chi_2 + \chi_3 \ ,
\end{equation} 
where
\begin{eqnarray*} 
\chi_1 &=& 
\phantom{+} t^2 \ \EE\left[ 
\frac{1}{n} \tr \left( 
{\bf DRH} \frac{{\bf Y} \widetilde{\bf D} \widetilde{\bf R}
{\bf Y}^*}{n} \right) 
\frac{1}{n} \tr\left( {\bf DHDH} \right) \right]\ , \\
\chi_2 &=& 
\ t^2 \ \EE\left[ 
% \frac{1}{n} \overbrace{\tr\left( {\bf DH} \right)}^{\circ} 
\overcirc\beta \, 
\frac{1}{n} \tr\left( 
{\bf DRHDH} \frac{{\bf Y} \widetilde{\bf D} \widetilde{\bf R}
{\bf Y}^*}{n} \right)
\right]\ , \\
\chi_3 &=& 
 \ \ \frac{1}{n} \tr\left( {\bf D}^2 {\bf R} 
\EE\left[ {\bf H} \right] \right) \ .
\end{eqnarray*}
Let us first deal with the terms $\chi_2$ and $\chi_3$. Cauchy-Schwarz
inequality together with Proposition \ref{PN-controle-1} and
Proposition \ref{grosse-proposition}-\eqref{variance-estimates} yield
$\chi_2={\mathcal O}(n^{-2})$. Proposition \ref{prop-E(H)-R} together with
Theorem \ref{theo-canonique} yield $\chi_3 = \gamma + {\cal
  O}(n^{-2})$. We now look at $\chi_1$. Due to Proposition \ref{grosse-proposition}-\eqref{variance-estimates}
and to Lemma \ref{lm-trDHDH}-\eqref{eq-var-trDHDH}, we have:
\begin{eqnarray*}
\chi_1 
&=& 
t^2 \ \EE\left[ 
\frac{1}{n} \tr \left( 
{\bf DRH} \frac{{\bf Y} \widetilde{\bf D} \widetilde{\bf R}
{\bf Y}^*}{n} \right) \right] 
\EE\left[ \frac{1}{n} \tr\left( {\bf DHDH} \right) \right] + 
{\cal O}\left(\frac{1}{n^2}\right)\ , \\
&\stackrel{(a)}{=}& 
t^2 \gamma \tilde\gamma
\EE\left[ \frac{1}{n} \tr\left( {\bf DHDH} \right) \right]  
+ 
{\cal O}\left(\frac{1}{n^2}\right)\ ,
\end{eqnarray*}
where $(a)$ follows from \eqref{approx-1} in Proposition \ref{grosse-proposition}.
It remains to plug the values obtained for $\chi_1$, $\chi_2$ and $\chi_3$ into (\ref{eq-trDHDH-1}) to obtain:
$$
(1 - t^2 \gamma \tilde\gamma)  \EE\left[ \frac{1}{n} \tr\left( {\bf DHDH} \right) \right] = \gamma 
+ {\cal O}(n^{-2})\ .
$$  
Recalling Proposition \ref{existence-variance}, we can divide by $(1 - t^2 \gamma \tilde\gamma)$ and 
obtain the desired result.
\end{proof}

We can now go back to the computation of $\EE \Psi({\bf Y})$. Let us give the main steps of the derivation. 
Expanding $\EE \Psi({\bf Y})$ yields: 
$$
\EE\left[ \frac 1n \tr\left( 
{\bf AHDH} \frac{ {\bf Y} {\bf B} {\bf Y}^* }
{n} \right) \right]
= 
\frac{1}{n^2} \sum_{p=1}^N \sum_{j=1}^n a_p b_j  
\EE\left[ \left[ {\bf HDHy}_j \right]_{p} \overline{Y_{pj}} \right] \ . 
$$  

We replace the summand $n^{-1} \EE\left[ \left[ {\bf HDHy}_j
  \right]_{p} \overline{Y_{pj}} \right]$ by the expression given by
(\ref{eq-HDHYpj)}). We then replace the term $\EE\left[\left[ {\bf
      HDH}\right]_{pp} \right]$ in \eqref{eq-HDHYpj)} by the expression given by
(\ref{eq-E(HDH)_pp}). We sum over $p$ and $j$ and notice afterwards that the
terms where $\overcirc{\beta}$ is involved are of order ${\mathcal
  O}(n^{-2})$. We therefore end up with:
\begin{eqnarray*} 
\EE\left[ \frac 1n \tr\left( 
{\bf AHDH} \frac{ {\bf Y} {\bf B} {\bf Y}^* }
{n} \right) \right]
&=& 
- t \ \EE\left[ \frac{1}{n} \tr\left( {\bf DHDH} \right)  \frac 1n \tr\left( 
{\bf AH} \frac{{\bf Y}\widetilde{\bf D} \widetilde{\bf R} {\bf BY}^*}{n}
\right)\right]  \\
& & 
+  \frac{t^2}{n} \tr\left( \widetilde{\bf D} \widetilde{\bf R}{\bf B} \right)
\EE\left[ \frac{1}{n} \tr\left( {\bf DHDH} \right)  \frac 1n \tr\left( 
{\bf AR} {\bf DH} \frac{{\bf Y}\widetilde{\bf D} \widetilde{\bf R} 
{\bf Y}^*}{n} \right)\right]  \\
& & 
+ \frac{1}{n} \tr\left( \widetilde{\bf D} \widetilde{\bf R} {\bf B} \right)
\frac{1}{n} \tr\left( {\bf A} {\bf D}^2 {\bf R}\, \mathbb{E}{\bf H} \right) 
+ {\cal O}\left(\frac{1}{n^2} \right) \ . 
\end{eqnarray*}
We first decorrelate by using the variance estimates in Proposition
\ref{grosse-proposition}-\eqref{variance-estimates} and Lemma \ref{lm-trDHDH}-\eqref{eq-var-trDHDH} and 
obtain:
\begin{eqnarray*} 
\EE\left[ \frac 1n \tr\left( 
{\bf AHDH} \frac{ {\bf Y} {\bf B} {\bf Y}^* }
{n} \right) \right]
&=& 
- t \ \EE\left[ \frac{1}{n} \tr\left( {\bf DHDH} \right) \right] 
\EE\left[ \frac 1n \tr\left( 
{\bf AH} \frac{{\bf Y}\widetilde{\bf D} \widetilde{\bf R} {\bf BY}^*}{n}
\right)\right]  \\
& & 
+ t^2 \frac{1}{n} \tr\left( \widetilde{\bf D} \widetilde{\bf R}{\bf B} \right)
\EE\left[ \frac{1}{n} \tr\left( {\bf DHDH} \right) \right] 
\EE\left[ \frac 1n \tr\left( 
{\bf AR} {\bf DH} \frac{{\bf Y}\widetilde{\bf D} \widetilde{\bf R} 
{\bf Y}^*}{n} \right)\right]  \\
& & 
+ \frac{1}{n} \tr\left( \widetilde{\bf D} \widetilde{\bf R} {\bf B} \right)
\frac{1}{n} \tr\left( {\bf A} {\bf D}^2 {\bf R}\,\mathbb{E}{\bf H} \right) 
+ {\cal O}\left(\frac{1}{n^2} \right) \ . 
\end{eqnarray*} 
It remains to apply Theorem \ref{theo-canonique}, Proposition \ref{grosse-proposition} 
and Lemma \ref{lm-trDHDH}-\eqref{eq-mean-trDHDH} to the terms in the right hand side
of the previous equality to conclude.

\bibliographystyle{IEEEbib}
\bibliography{BSTLabbrev,bibli}

\def\cprime{$'$}
\begin{thebibliography}{10}

\bibitem{tel-95}
I.E. Telatar,
\newblock ``{Capacity of Multi-Antenna Gaussian Channels},''
\newblock {\em \emph{published in} European Transactions on
  Telecommunications}, vol. 10, no. 6, pp. 585--595, Nov/Dec 1999,
\newblock Technical Memorandum, Bell Laboratories, Lucent Technologies, October
  1995.

\bibitem{fos-gan-98}
Foschini G.J. and M.J. Gans,
\newblock ``{On Limits of Wireless Communications in a Fading Environment when
  Using Multiple Antennas},''
\newblock {\em Wireless Personal Communications}, vol. 6, no. 3, pp. 311--335,
  Mar. 1998.

\bibitem{mar-pas-ms67}
V.A. Marchenko and L.A. Pastur,
\newblock ``{Distribution of Eigenvalues for Some Sets of Random Matrices},''
\newblock {\em Math. USSR -- Sbornik}, vol. 1, no. 4, pp. 457--483, 1967.

\bibitem{mue-it02}
R.R. M\"uller,
\newblock ``{A Random Matrix Model of Communication Via Antenna Arrays},''
\newblock {\em IEEE Trans. on Information Theory}, vol. 48, no. 9, pp.
  2495--2506, Sept. 2002.

\bibitem{chu-tse-kah-val-it02}
C.-N. Chuah, D.N.C. Tse, J.M. Kahn, and R.A. Valenzuela,
\newblock ``{Capacity Scaling in MIMO Wireless Systems Under Correlated
  Fading},''
\newblock {\em IEEE Trans. on Information Theory}, vol. 48, no. 3, pp.
  637--650, Mar. 2002.

\bibitem{mes-fon-pag-jsac03}
X.~Mestre, J.R. Fonollosa, and A.~Pag\`es-Zamora,
\newblock ``{Capacity of MIMO channels: Asymptotic Evaluation Under Correlated
  Fading},''
\newblock {\em IEEE Journal on Selected Areas in Communications}, vol. 21, no.
  5, pp. 829--838, June 2003.

\bibitem{mou-sim-sen-it03}
A.L. Moustakas, S.H. Simon, and A.M. Sengupta,
\newblock ``{MIMO Capacity Through Correlated Channels in the Presence of
  Correlated Interferers and Noise: A (Not So) Large N Analysis },''
\newblock {\em IEEE Trans. on Information Theory}, vol. 49, no. 10, pp.
  2545--2561, Oct. 2003.

\bibitem{tul-loz-ver-it05}
A.M. Tulino, A.~Lozano, and S.~Verd{\'u},
\newblock ``{Impact of Antenna Correlation on the Capacity of Multiantenna
  Channels},''
\newblock {\em IEEE Trans. on Information Theory}, vol. 51, no. 7, pp.
  2491--2509, July 2005.

\bibitem{hac-lou-naj-(sub)aap05}
W.~Hachem, Ph. Loubaton, and J.~Najim,
\newblock ``{Deterministic Equivalents for Certain Functionals of Large Random
  Matrices},''
\newblock {\em \emph{accepted for publication in} Annals of Applied
  Probability}, 2006,
\newblock \texttt{arXiv:math.PR/0507172}.

\bibitem{HLN05}
W.~Hachem, P.~Loubaton, and J.~Najim,
\newblock ``The empirical eigenvalue distribution of a gram matrix: from
  independence to stationarity,''
\newblock {\em Markov Process. Related Fields}, vol. 11, no. 4, pp. 629--648,
  2005.

\bibitem{shi-fos-gan-kah-tcom00}
Shiu D.-S., G.J. Foschini, M.J. Gans, and J.M. Kahn,
\newblock ``{Fading Correlation and its Effect on the Capacity of Multielement
  Antenna Systems},''
\newblock {\em IEEE Trans. on Communications}, vol. 48, no. 3, pp. 502--513,
  Mar. 2000.

\bibitem{ker-sch-perd-mog-fre-jsac02}
J.P. Kermoal, L.~Schumacher, K.I. Pedersen, P.E. Mogensen, and F.~Frederiken,
\newblock ``{A Stochastic MIMO Radio Channel Model with Experimental
  Validation},''
\newblock {\em IEEE Journal on Selected Areas in Communications}, vol. 20, no.
  6, pp. 1211--1225, 2002.

\bibitem{AndZei06}
G.~W. Anderson and O.~Zeitouni,
\newblock ``A {CLT} for a band matrix model,''
\newblock {\em Probab. Theory Related Fields}, vol. 134, no. 2, pp. 283--338,
  2006.

\bibitem{BaiSil04}
Z.~D. Bai and J.~W. Silverstein,
\newblock ``C{LT} for linear spectral statistics of large-dimensional sample
  covariance matrices,''
\newblock {\em Ann. Probab.}, vol. 32, no. 1A, pp. 553--605, 2004.

\bibitem{BouKho98}
A.~Boutet~de Monvel and A.~Khorunzhy,
\newblock ``Limit theorems for random matrices,''
\newblock {\em Markov Process. Related Fields}, vol. 4, no. 2, pp. 175--197,
  1998.

\bibitem{KhoPas93}
A.~M. Khorunzhy and L.~A. Pastur,
\newblock ``Limits of infinite interaction radius, dimensionality and the
  number of components for random operators with off-diagonal randomness,''
\newblock {\em Comm. Math. Phys.}, vol. 153, no. 3, pp. 605--646, 1993.

\bibitem{KPV95}
L.~A. Pastur, A.~M. Khorunzhi{\u\i}, and V.~Yu. Vasil{\cprime}chuk,
\newblock ``On an asymptotic property of the spectrum of the sum of
  one-dimensional independent random operators,''
\newblock {\em Dopov. Nats. Akad. Nauk Ukra\"\i ni}, , no. 2, pp. 27--30, 1995.

\bibitem{pas-umj05}
L.A. Pastur,
\newblock ``{A Simple Approach to the Global Regime of Gaussian Ensembles of
  Random Matrices},''
\newblock {\em Ukrainian Mathematical Journal}, vol. 57, no. 6, pp. 936--966,
  June 2005.

\bibitem{Che82}
L.~H.~Y. Chen,
\newblock ``An inequality for the multivariate normal distribution,''
\newblock {\em J. Multivariate Anal.}, vol. 12, no. 2, pp. 306--315, 1982.

\bibitem{HPS98}
C.~Houdr{\'e}, V.~P{\'e}rez-Abreu, and D.~Surgailis,
\newblock ``Interpolation, correlation identities, and inequalities for
  infinitely divisible variables,''
\newblock {\em J. Fourier Anal. Appl.}, vol. 4, no. 6, pp. 651--668, 1998.

\bibitem{hor-joh-livre94}
R.~Horn and C.~Johnson,
\newblock {\em {Matrix Analysis}},
\newblock Cambridge Univ. Press, 1994.

\bibitem{GliJaf87}
J.~Glimm and A.~Jaffe,
\newblock {\em Quantum physics},
\newblock Springer-Verlag, New York, second edition, 1987,
\newblock A functional integral point of view.

\end{thebibliography}

\end{document}